\documentclass[conference]{IEEEtran}

\usepackage[T1]{fontenc}
\usepackage{graphicx}
\usepackage{url}
\usepackage{color}
\usepackage{colortbl}
\usepackage{amsmath}
\usepackage{xspace}
\usepackage{epsfig}
\usepackage{subfigure}
\usepackage{balance}
\usepackage{amsfonts}
\usepackage{ntheorem}
\usepackage{nccmath}
\usepackage[normalem]{ulem}
\usepackage{balance}
\usepackage{cite}
\usepackage{subfigure}
\usepackage{float}
\usepackage[dvipsnames]{xcolor}

\usepackage{tabu}
\usepackage{paralist}
\usepackage{diagbox}
\usepackage{booktabs}
\usepackage{multirow}
\usepackage[ruled,linesnumbered,vlined]{algorithm2e}

\theoremseparator{.}
\theoremheaderfont{\scshape}

\newtheorem{example}{Example}

\newtheorem{remark}{Remark}
\newtheorem{observation}{Observation}

\newtheorem{drule}{Rule}

\newcommand{\stitle}[1]{\noindent{\bf #1}}

\newcommand{\reffig}[1]{Figure~\ref{fig:#1}}
\newcommand{\refsec}[1]{Section~\ref{sec:#1}}
\newcommand{\refsubsec}[1]{Section~\ref{subsec:#1}}
\newcommand{\reftab}[1]{Table~\ref{tab:#1}}
\newcommand{\refalg}[1]{Algorithm~\ref{alg:#1}}
\newcommand{\refeq}[1]{Equation~\ref{eq:#1}}

\newcommand{\refex}[1]{Example~\ref{ex:#1}}
\newcommand{\refob}[1]{Observation~\ref{ob:#1}}

\newcommand{\refrul}[1]{Rule~\ref{rul:#1}}

\newcommand{\kw}[1]{{\ensuremath {\mathsf{#1}}}\xspace}

\newcommand{\Roam}[1]{\uppercase\expandafter{\romannumeral #1}}
\newcommand{\roam}[1]{{\romannumeral #1}}

\newcommand{\norm}[1]{\left\lVert#1\right\rVert}

\newcommand{\vp}{\textbf{p}}
\newcommand{\vq}{\textbf{q}}
\newcommand{\loss}{{\cal L}}

\newcommand{\algnaive}{HSGD\xspace}
\newcommand{\algopt}{HSGD*\xspace}

\newcommand{\gpunum}{n_g}
\newcommand{\cpunum}{n_c}

\newcommand{\algcpu}{CPU-Only\xspace}
\newcommand{\alggpu}{GPU-Only\xspace}

\newcommand{\ml}{MovieLens\xspace}
\newcommand{\nf}{Netflix\xspace}
\newcommand{\ro}{R1\xspace}
\newcommand{\yh}{Yahoo!Music\xspace}

\begin{document}
\title{Efficient Matrix Factorization on Heterogeneous CPU-GPU Systems}

\author{
	Yuanhang Yu$^{\natural}$, Dong Wen$^{\natural}$, Ying Zhang$^{\natural}$, Xiaoyang Wang$^{\S}$, Wenjie Zhang$^{\dag}$, and Xuemin Lin$^{\dag}$ \vspace{1mm}\\
	\fontsize{10}{10}\selectfont\itshape
	$^{\natural}$AAII, University of Technology Sydney, Australia\\
	$^{\S}$Zhejiang Gongshang University, China\\
	$^{\dag}$The University of New South Wales, Australia\\
	\fontsize{9}{9}\selectfont\ttfamily\upshape
	$^{\natural}$yuanhang.yu@student.uts.edu.au; \{dong.wen, ying.zhang\}@uts.edu.au;\\
	$^{\S}$xiaoyangw@zjgsu.edu.cn; $^{\dag}$\{zhangw, lxue\}@cse.unsw.edu.au;
}

\maketitle

\begin{abstract}
Matrix Factorization (MF) has been widely applied in machine learning and data mining. A large number of algorithms have been studied to factorize matrices. Among them, stochastic gradient descent (SGD) is a commonly used method. Heterogeneous systems with multi-core CPUs and GPUs have become more and more promising recently due to the prevalence of GPUs in general-purpose data-parallel applications. Due to the large computational cost of MF, we aim to improve the efficiency of SGD-based MF computation by utilizing the massive parallel processing power of heterogeneous multiprocessors. The main challenge in parallel SGD algorithms on heterogeneous CPU-GPU systems lies in the granularity of the matrix division and the strategy to assign tasks. We design a novel strategy to divide the matrix into a set of blocks by considering two aspects. First, we observe that the matrix should be divided nonuniformly, and relatively large blocks should be assigned to GPUs to saturate the computing power of GPUs. In addition to exploiting the characteristics of hardware, the workloads assigned to two types of hardware should be balanced. Aiming at the final division strategy, we design a cost model tailored for our problem to accurately estimate the performance of hardware on different data sizes. A dynamic scheduling policy is also used to further balance workloads in practice. Extensive experiments show that our proposed algorithm achieves high efficiency with a high quality of training quality.
\end{abstract}

\section{Introduction}
\label{sec:introduction}
As a common technique in machine learning and data mining, matrix factorization (MF)
has been widely applied in many areas, such as recommendation system~\cite{lian2014geomf,he2016fast}, social network analysis~\cite{li2015predicting,al2017unveiling}, web mining~\cite{qiu2018network}, word embedding~\cite{pennington2014glove}, and graph embedding~\cite{DBLP:journals/tkde/CuiWPZ19}.

Given a sparse matrix $R$ with $m$ rows and $n$ columns, the aim of MF is to decompose $R$ into two dense matrices $P(m \times k)$ and $Q(k \times n)$, which satisfies the following condition:

\vspace{-0.5em}
\begin{center}
$R \approx P \times Q$
\end{center}
\vspace{-0.5em}

\noindent Here, the dimension $k$ is far less than $m$ and $n$ so that the original sparse data
can be represented well by new dense data with low dimensionality. 

\stitle{Stochastic Gradient Descent.} The existing literature has proposed three main approaches to solve MF, namely, alternating least squares (ALS), coordinate descent (CD), and stochastic gradient descent (SGD). Among them, SGD-based algorithms have received the most attention, due to their algorithmic simplicity and effectiveness. 
SGD algorithms execute several iterations to make the training result convergent. The number of iterations can be a parameter specified by users. In each iteration, the gradient of every element in the sparse matrix $R$ is computed and the corresponding vectors in the result matrices are updated. 
We focus on the SGD-based MF algorithms in this paper and use MF to denote the SGD-based MF whenever there is no ambiguity. 

In the literature, several parallel SGD approaches have been proposed on different system settings such as GPUs (e.g., CuMF\_SGD~\cite{xie2017cumf_sgd}), multi-core CPUs (e.g., FPSGD~\cite{zhuang2013fast}), and distributed systems (e.g., NOMAD~\cite{yun2014nomad}). 
To parallelize the computation in SGD, the main idea of these methods is to uniformly divide the sparse matrix $R$ into a set of blocks. In each iteration, a set of mutually independent blocks are assigned to different working units (e.g., threads, nodes, or GPUs) and updated by these working units. Here, two blocks are independent of each other if they do not share the same row and the same column. This strategy avoids the writing conflict in $P$ and $Q$ since the elements in the same row (resp. column) of $R$ will update the same row (resp. column) of $P$ (resp. $Q$). Several optimizations are also studied for their specific system settings, and more related works of MF are introduced in \refsec{existingwork}. 

\begin{figure}[t!]
\begin{center}
\includegraphics[width=0.8\columnwidth,natwidth=620,natheight=233]{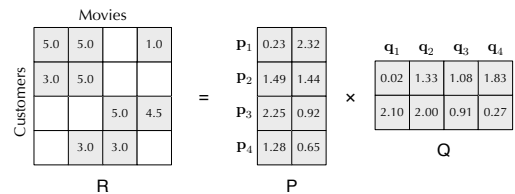}
\vspace{-1em}
\caption{A rating matrix $R$ and a corresponding matrix factorization}
\label{fig:mf_exp}
\end{center}
\vspace{-0.5em}
\end{figure}

Nowadays, GPU becomes very popular for running general-purpose data-parallel applications. Consequently, heterogeneous systems with multi-core CPUs and GPUs become more and more promising for many general tasks. The idea of combing CPU and GPU not only speeds up the task execution but also effectively utilizes the computing resources available in heterogeneous systems. By contrast, to meet performance requirements, the CPU-only or GPU-only systems are usually over-provisioned. This leads that their average utilization remains low. For example, after allocating the task to a GPU (i.e., starting the kernel), the CPU stays idle. Similarly, for applications where the GPU memory bandwidth is a major bottleneck, the massive computation resources of GPUs remain underutilized. If the resources that were originally idle in the system are utilized effectively, the running time of the program will be reduced. This has motivated a significant amount of research on heterogeneous computing techniques \cite{luk2009qilin,pandit2014fluidic,stehle2017memory}. To our best knowledge, there is no existing work on the MF task on heterogeneous CPU-GPU systems, and it is desirable to develop an efficient MF algorithm for this system setting. In this paper, we mainly study the task division and the scheduling strategy for SGD on heterogeneous CPU-GPU systems. Our techniques do not closely depend on any specific algorithm tailored for a single thread of the CPU or a GPU.

Due to the close relation between threads in a GPU, a straightforward method to combine GPUs and CPUs is to consider the GPU as a CPU thread. Then, we can naturally adopt the state-of-the-art shared memory algorithm FPSGD \cite{zhuang2013fast} in our problem. We call this method \algnaive. Despite the success of \algnaive, there are still opportunities for further improvement. First, based on \cite{zhuang2013fast}, we derive a matrix division rule such that the number of blocks in the matrix should be at least $(\cpunum+\gpunum+1) \times (\cpunum+\gpunum)$, where $\cpunum$ and $\gpunum$ are the number of CPU cores and GPUs, respectively. We observe that in \algnaive, even with the smallest number of blocks, the size of a block derived by the uniform division is not large enough to saturate the processing power of GPUs (\refob{small_gpu}). This may limit the overall efficiency of \algnaive. Second, the computing power of a GPU is normally much stronger than that of a single CPU thread. Given that blocks sharing the same row or column cannot be processed in parallel, the update frequency of blocks processed by GPU can be extremely high due to the powerful computing power of GPU. This phenomenon may lead to a weak training quality (\refex{poor_training_result}).

\stitle{Our Approach.} To overcome the drawbacks in \algnaive, we propose a nonuniform matrix division strategy for the parallel SGD algorithm. 
Specifically, we first design a novel cost model to evaluate the performance of hardware. Existing works on estimating the working efficiency on heterogeneous CPU-GPU systems are neither available in our setting nor accurate enough in the MF problem. Our cost model is tailored for the parallel SGD algorithm and can accurately estimate the performance of GPUs when the block is relatively small. Based on the cost model, we can initially divide the matrix into two parts which are assigned to GPUs and CPUs, respectively. Each part is divided into a set of blocks to avoid conflicts in MF computation.
We also adopt a dynamic scheduling policy which allows the dominant computation resource to use work-stealing mechanism\cite{blumofe1999scheduling} and process blocks originally assigned to the other resource. This optimization further balances workloads by filling the gap between the estimation by our cost model and the practical performance. When the dynamic scheduling policy is activated, the sparse matrix $R$ is divided into more blocks for avoiding conflicts.

\stitle{Contributions.} We summarize main contributions as follows.

\begin{itemize}
\item \textit{A parallel SGD algorithm on heterogeneous CPU-GPU systems.} To our best knowledge, this is the first work to efficiently tackle Matrix Factorization on heterogeneous CPU-GPU systems.

\item \textit{A nonuniform matrix division strategy to improve efficiency.} Compared with a straightforward method to uniformly divide the matrix, our division strategy fully utilizes the strong processing power of GPUs.

\item \textit{A novel cost model to estimate the performance of GPUs.} We balance workloads using a cost model specifically designed for our problem. The cost model of the GPU part considers both data transfer and kernel execution.

\item \textit{Extensive experiments on four benchmark datasets.} The experiments demonstrate that our proposed framework can effectively utilize the resources in the heterogeneous CPU-GPU systems and show the superior performance compared to the competitors.
\end{itemize}

\stitle{Organization.} The rest of the paper is organized as follows. \refsec{pre} introduces background information including MF and GPU architecture. \refsec{existingwork} reviews related works. \refsec{overview} discusses the motivation and proposes an overall framework. \refsec{model} proposes the details of our cost model. \refsec{task} introduces the dynamic scheduling strategy and the final matrix division strategy. \refsec{exp} reports the experimental results, and \refsec{conclusion} concludes the paper.

\section{Preliminary}
\label{sec:pre}

\subsection{Matrix Factorization}

We consider a user-item rating matrix $R \in \mathbb{R}^{m \times n}$ where $m$ and $n$ are the number of rows and the number of columns of the matrix, respectively. For each $1 \le u \le m$ and $1 \le v \le n$, $r_{u,v}$ is the rating from the user $u$ to the item $v$. $R$ is normally sparse since not every element in $R$ is explicitly reported. Matrix factorization aims to represent the matrix $R$ as a dot product between two dense matrices $P \in \mathbb{R}^{m \times k}$ and $Q \in \mathbb{R}^{k \times n}$, where $k$ is the number of latent factors. A mathematical representation of the matrix factorization is shown as follows.

\begin{equation}
\label{eq:approximate}
r_{u,v} \approx \vp_{u} \vq_{v}
\end{equation}

In \refeq{approximate}, $\vp_{u}$ is the $u$-th row vector of $P$, and $\vq_{v}$ is the $v$-th column vector of $Q$. Matrix factorization achieves \refeq{approximate} by minimizing the following loss function.

\begin{equation}
\label{eq:sgd}
\loss=\mathop{\sum}\limits_{(u,v) \in R}(r_{u,v}-\vp_{u}\vq_{v})^2+\lambda_{P}\norm{\vp_{u}}_{F}^{2}+\lambda_{Q}\norm{\vq_{v}}_{F}^{2}
\end{equation}
\vspace{-0.6em}

In \refeq{sgd}, $(r_{u,v}-\vp_{u}\vq_{v})^2$ measures the gap between $r_{u,v}$ and estimated value $\vp_{u}\vq_{v}$. $\lambda_{P}\norm{\textbf{p}_{u}}_{F}^{2}$ and $\lambda_{Q}\norm{\textbf{q}_{v}}_{F}^{2}$ are used to avoid over-fitting. $\norm{.}_{F}^{2}$ computes the Frobenius norm\footnote{\url{https://en.wikipedia.org/wiki/Matrix_norm\#Frobenius_norm}} of a vector. $\lambda_{P}$ and $\lambda_{Q}$ are regularization coefficients.

\vspace{-0.5em}
\begin{example}
We give an example of the matrix factorization in \reffig{mf_exp}. The rating matrix $R$ contains nine ratings for four movies given by four customers. The number of latent factors $k$ is $2$. We have $r_{1,2}=5.0$ which means that $u_{1}$ gives a rating $5.0$ to $v_{2}$. The results of $R$'s matrix factorization are shown on the right of $R$. $P$ is a user preference matrix, and $Q$ is a movie feature matrix. The vector $\vp_{1} (0.23, 2.32)$ is the preference of the user $u_{1}$, and $\vq_{2} (1.33, 2.00)^{T}$ is the feature of the movie $v_{2}$. The estimated value of $\textbf{p}_{1}\textbf{q}_{2}$ is $4.9459$, which is close to $R_{1,2}$ $5.0$.
\end{example}

\subsection{Stochastic Gradient Descent for Matrix Factorization}

It is time-consuming to calculate the loss of the whole matrix $R$ when using the loss function of \refeq{sgd}, especially when $R$ contains billions of items. Several works have been done to minimize \refeq{sgd} and improve the efficiency of matrix factorization \cite{funk2006netflix,koren2009matrix,yu2012scalable}. In this paper, we follow a prevalent algorithm called \textit{stochastic gradient descent} (SGD) \cite{funk2006netflix} among them, and ideas of several other algorithms are introduced in \refsec{existingwork}.

SGD executes several iterations. The number of iterations can be specified by users. Instead of straightforwardly applying the gradient descent to minimize \refeq{sgd} in each iteration, SGD computes the gradient of every element $r_{u,v}$ in $R$ and updates the corresponding vectors in the result matrices. Consequently, the original loss function in \refeq{sgd} is reduced to the following equation.

\vspace{-1em}
\begin{center}
\begin{equation}
\label{eq:simple_sgd}
\loss = (r_{u,v}-\textbf{p}_{u}\textbf{q}_{v})^2+\lambda_{P}\textbf{p}_{u}\textbf{p}_{u}^{T}+\lambda_{Q}\textbf{q}_{v}^{T}\textbf{q}_{v}
\end{equation}
\end{center}

The gradient of \refeq{simple_sgd} is represented as follows.

\vspace{-1em}
\begin{center}
\begin{equation}
\label{eq:gradient_p}
\frac{\partial L}{\partial \textbf{p}_{u}}=2(\textbf{p}_{u}\textbf{q}_{v}-r_{u,v})\textbf{q}_{v}^T+2\lambda_{P}\textbf{p}_{u}
\end{equation}
\end{center}
\vspace{-1em}

\begin{center}
\begin{equation}
\label{eq:gradient_q}
\frac{\partial L}{\partial \textbf{q}_{v}}=2(\textbf{p}_{u}\textbf{q}_{v}-r_{u,v})\textbf{p}_{u}^T+2\lambda_{Q}\textbf{q}_{v}
\end{equation}
\end{center}
\vspace{-0.5em}

Based on the gradient (\refeq{gradient_p} and \refeq{gradient_q}), we train the model iteratively, and the value of the loss function decreases until it is convergent. \refeq{gradient_descent} shows this process where $\gamma$ is the learning rate.

\vspace{-1em}
\begin{center}
\begin{equation}
\label{eq:gradient_descent}
\textbf{p}_{u} \gets \textbf{p}_{u}-\gamma\frac{\partial L}{\partial \textbf{p}_{u}}~~~~~\textbf{q}_{v} \gets \textbf{q}_{v}-\gamma\frac{\partial L}{\partial \textbf{q}_{v}} 
\end{equation}
\end{center}

\begin{algorithm}[t!]
\caption{SGD}
\label{alg:SGD-based_MF_Algorithm}
\KwIn{$R_{m\times n}, k, \lambda_{P}, \lambda_{Q}, \gamma, t$}
\KwOut{$P_{m \times k}, Q_{k \times n}$}

\tcp{Data preprocessing phase}

$\kw{init\_model}(P_{m\times k}, Q_{k\times n})$\;

\tcp{Training phase}
\ForEach{$iteration \gets 1$ \KwTo $t$}{
  \ForEach{$r_{u, v} \in R_{m \times n}$}{

    $e_{u, v} \gets r_{u, v} - \vp_{u}\vq_{v}$\;
    $\vp_{u} \gets \vp_{u} + \gamma(e_{u, v}\vq_{v}^T - \lambda_{P}\vp_{u})$\;
    $\vq_{v} \gets \vq_{v} + \gamma(e_{u, v}\vp_{u}^T - \lambda_{Q}\vq_{v})$\;
  }
}

\tcp{Data post-processing phase}
$\kw{save\_model}(P_{m\times k}, Q_{k\times n})$\;
\end{algorithm}

We present the pseudocode of SGD in \refalg{SGD-based_MF_Algorithm}. There are several input parameters. $R_{m \times n}$ is a sparse rating matrix stored in the form of triadic tuple. $k$ is the number of latent factors. $\gamma$ is the learning rate. $\lambda$ is the regularization parameter. $t$ is the number of iterations. \refalg{SGD-based_MF_Algorithm} outputs two feature matrices $P_{m \times k}$ and $Q_{k \times n}$.
The algorithm consists of three phases: data preprocessing, SGD training, and data post-processing.
The data preprocessing phase initializes two resulting matrices $P$ and $Q$ with values generated randomly.
In the training phase, in each iteration (line 3-6), every element is picked to decrease the value of loss function by using \refeq{gradient_descent}. The SGD training terminates when the given number of iterations is reached (line 2) or the model converges.
The feature matrices are stored after training.

\stitle{Problem Definition.} In this paper, we aim to develop an efficient SGD-based MF algorithm in heterogeneous CPU-GPU systems. 

\vspace{-0.5em}
\begin{remark}
Our research mainly focuses on the scheduling strategy for the task division and assignment between CPUs and GPUs. Our proposed techniques will not closely depend on any specific GPU or GPU SGD-based MF algorithms. 
\end{remark}

\subsection{Characteristics of GPUs}
\label{subsec:GPU_knowledge}

We introduce several basic architecture characteristics of GPUs. More details can be found in \cite{jia2018dissecting}.

\stitle{Execution Model.} GPU manages thousands of threads in a three-layer hierarchy, which includes grids, blocks, and warps. A block contains dozens of warps, and multiple blocks are organized into a grid. A warp is the smallest execution unit from the view of hardware and contains 32 consecutive threads. Instructions of a GPU program (also named kernel) are executed in a way called SIMT (Single Instruction Multiple Thread). SIMT means that all threads in a warp must execute the same instruction on their own data at the same time, while the state of threads in a warp can be different (active or inactive) so that they can go through different execution paths. Therefore, GPU threads are not as flexible as CPU threads. 

\stitle{Memory Hierarchy.} GPU has a complicated memory hierarchy. We introduce several commonly used concepts. \textit{Global memory} is the largest memory, whose access speed is the slowest. Data in global memory can be accessed by all threads. \textit{Shared memory} is on-chip and a relatively scarce resource. It can be accessed by threads from the same thread block since GPU promises that threads in the same block are performed on the same processor. \textit{Register} is used to store local variables in kernel, whose access speed is the fastest. A register is only directly accessed by the thread it belongs to. Threads can access local variables of other threads in the same warp. 

\section{Related Works}
\label{sec:existingwork}
In this section, we introduce the details of FPSGD \cite{zhuang2013fast} and CuMF\_SGD \cite{xie2017cumf_sgd}, which are related to our problem. Several other related works are briefly summarized.

\subsection{Multi Core MF Algorithms}
\label{subsec:cpu_algorithm}

There are several algorithms extending the ideas of SGD in shared memory (multi CPU cores) systems. Hogwild~\cite{recht2011hogwild} adopts a lock-free update strategy to compute MF in parallel. DSGD~\cite{gemulla2011large} performs a matrix-blocking partition to avoid conflicts. Several variants of DSGD \cite{zhuang2013fast,oh2015fast,li2013sparkler} also follow this property. FPSGD \cite{zhuang2013fast} is the state-of-the-art in this type. We introduce more details of FPSGD as follows.

\stitle{FPSGD.} FPSGD uniformly divides the rating matrix $R$ into a set of sub-matrices (called blocks). Two blocks are independent of each other if they share neither any common column nor any common row of the rating matrix. Otherwise, we say they conflict. FPSGD applies the SGD algorithm to a set of independent blocks simultaneously and repeats this step until the number of processed blocks reaches a predefined value. In each iteration, all selected blocks must be independent since two blocks in the same row (resp. column) update the same area of $P$ (resp. $Q$) and cannot be processed in parallel.

\begin{figure}[t!]
\begin{center}
\includegraphics[width=0.3\columnwidth]{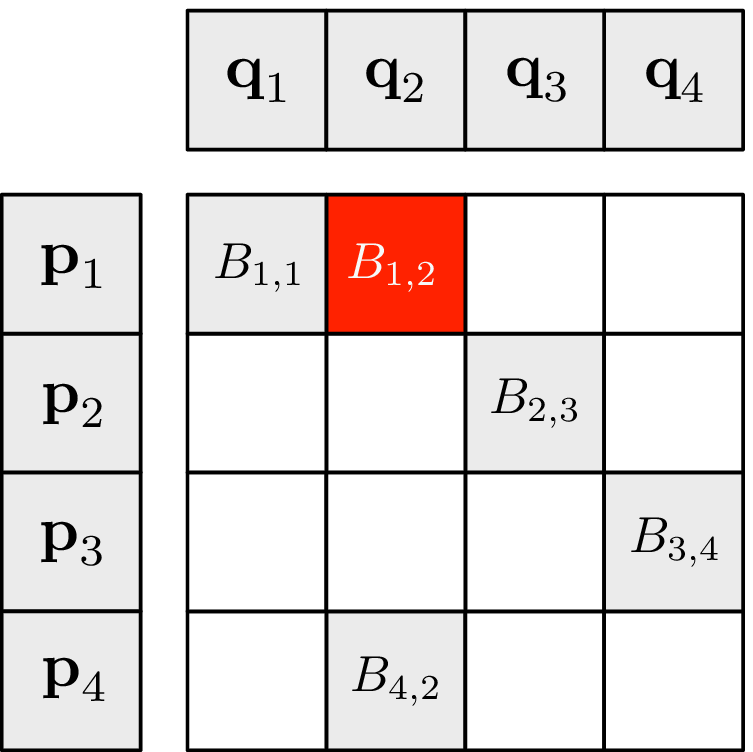}
\vspace{-0.5em}
\caption{An example of independent and conflicting blocks in $4 \times 4$ grid matrix}
\label{fig:fpsgd_exp}
\end{center}
\vspace{-0.5em}
\end{figure}

\begin{example}
We explain FPSGD in \reffig{fpsgd_exp}. The rating matrix $R$ is divided into $4 \times 4$ blocks. The computation on $B_{1,1}$ updates vectors $\vp_1$ and $\vq_1$, and the computation on $B_{2,3}$ updates vectors $\vp_2$ and $\vq_3$. They update vectors in different regions, which means that the computation on $B_{1,1}$ and $B_{2,3}$ can be performed at the same time. We can see that $B_{1,1}, B_{2,3}, B_{3,4}$ and $B_{4,2}$ are mutually independent. However, $B_{1,1}$ and $B_{1,2}$ are not independent of each other since they both update vectors in $\textbf{p}_1$, and this leads to a conflict.
\end{example}

\subsection{GPU Algorithms}
\label{subsec:gpu_algorithm}
GPUSGD~\cite{jin2016gpusgd} proposes matrix blocking-based MF solution on GPUs. BIDMach~\cite{canny2013bidmach} supports SGD-based MF and uses GPUs as accelerators. CuMF\_SGD\cite{xie2017cumf_sgd} is the state-of-the-art.

\stitle{CUMF\_SGD.} CuMF\_SGD designs a high-performance GPU kernel by fully exploiting GPU hardware characteristics, which includes warp shuffle, memory coalescing, register usage, and half-precision.
CuMF\_SGD fully utilizes the CPU-GPU memory transfer bandwidth by simultaneously performing the memory transfer and the computation. Specifically, when scheduling blocks, the algorithm randomly selects multiple consecutive blocks at a time instead of only one independent block for the GPU. The strategy reduces the overhead of CPU-GPU memory transfer. CuMF\_SGD uses CUDA streams to achieve this strategy. A stream contains a list of GPU commands that are executed in serial. Commands in different streams are executed in parallel if hardware resources permit. CuMF\_SGD kernel uses three streams to manage the CPU-GPU memory transfer, the GPU kernel, and the GPU-CPU memory transfer, respectively.

\subsection{Other Related Works}
\label{subsec:other_work}

\stitle{Other SGD-based MF Algorithms.} MF has been extensively investigated in the literature and a large number of algorithms were proposed for this problem. In addition to algorithms on multi-core and GPU architecture mentioned before, Parallel SGD solutions have been discussed in distributed systems~\cite{li2013sparkler}~\cite{yun2014nomad} and parameter server~\cite{zhong2016scaling}.

\stitle{Other MF Algorithms.} In addition to SGD-based algorithms, other methods like Coordinate Descent (CD) \cite{yu2012scalable} and Alternate Least Square (ALS) \cite{koren2009matrix} are also proposed to compute MF. They use different update rules in each iteration. Specifically, ALS \cite{koren2009matrix} updates one of the result matrices once by fixing the other. Then, it updates the other result matrix in the same way. An iteration is completed when two result matrices are both updated. CD \cite{yu2012scalable} updates one element in a result matrix once by fixing all other elements in two result matrices. An iteration is completed when all elements in result matrices are updated by the same update rule.

\section{Our Approach}
\label{sec:overview}
In this section, we first give a straightforward method and show its drawbacks. Then, we briefly introduce scheduling algorithms for heterogeneous systems and propose our method to balance workloads. Finally, an overview of our improved algorithm is provided.

\vspace{-0.5em}
\subsection{A Straightforward Method}
\label{subsec:straightforward_method}
\vspace{-0.2em}

A straightforward idea to utilize both CPU and GPU resources is to treat a GPU kernel as a worker thread. Based on this idea, we can adapt FPSGD \cite{zhuang2013fast} by regarding a GPU as an additional worker thread. Similarly, to avoid the conflict and obtain good training quality, a worker thread receives a new block satisfying the two criteria in \refsubsec{cpu_algorithm} once it finishes processing a block. We apply the FPSGD and CuMF\_SGD algorithms to process a matrix block on a CPU thread and a GPU kernel, respectively. This straightforward algorithm shown above can be called \algnaive.

Let $\cpunum$ and $\gpunum$ be the number of CPU threads and GPUs. Following the matrix-division rule in \cite{zhuang2013fast}, the number of blocks should be at least $(\cpunum + \gpunum +1) \times (\cpunum + \gpunum + 1)$. We refine this formula and give a more precise matrix-division rule.

\vspace{-0.2em}
\begin{drule}
\label{rul:rule1}
Given an input rating matrix $R$, $\cpunum$ CPU threads, and $\gpunum$ GPUs, $R$ should be divided into at least $(\cpunum + \gpunum +1) \times (\cpunum + \gpunum)$ blocks.
\end{drule}
\vspace{-0.2em}

We explain the rationale of this rule. When the number of blocks is less than $(\cpunum + \gpunum +1)\times(\cpunum + \gpunum)$, every thread is always assigned the same block. As a result, only several specific blocks are updated during the algorithm. Obviously, this will lead to a terrible training result. Even worse, the algorithm cannot fully exploit all worker threads. For example, in \reffig{fpsgd_exp}, assume that there are $4$ threads. The block $B_{1,1}$ is assigned to thread 1, and the rest gray blocks are assigned to other threads. When thread 1 finishes processing $B_{1,1}$, the rest gray blocks may still be occupied. To avoid conflicts, thread 1 has to continually process $B_{1,1}$. By contrast, when the block number increases to $(\cpunum + \gpunum +1)\times(\cpunum + \gpunum)$, thread 1 can always locate a spare row or column which is not occupied by other blocks.




\vspace{-0.5em}
\subsection{Motivation}
\label{subsec:Motivation}
\vspace{-0.2em}

Even though the \algnaive successfully combines CPU and GPU resources, we have several observations which can help (\roam{1}) improve the working efficiency of GPUs and (\roam{2}) balance workloads for different hardware. 


\stitle{Exploiting Hardware Characteristics.}
GPUs and CPUs have different features. We have two observations as follows.

\begin{observation}
\label{ob:small_gpu}
In the context of MF, small blocks cannot saturate the GPU computing power.
\end{observation}
\vspace{-0.5em}

\begin{figure}[t!]
\begin{center}
\begin{tabular}[t]{c}
\hspace{-1em}
\subfigure[GPU]{
  \includegraphics[width=0.5\columnwidth]{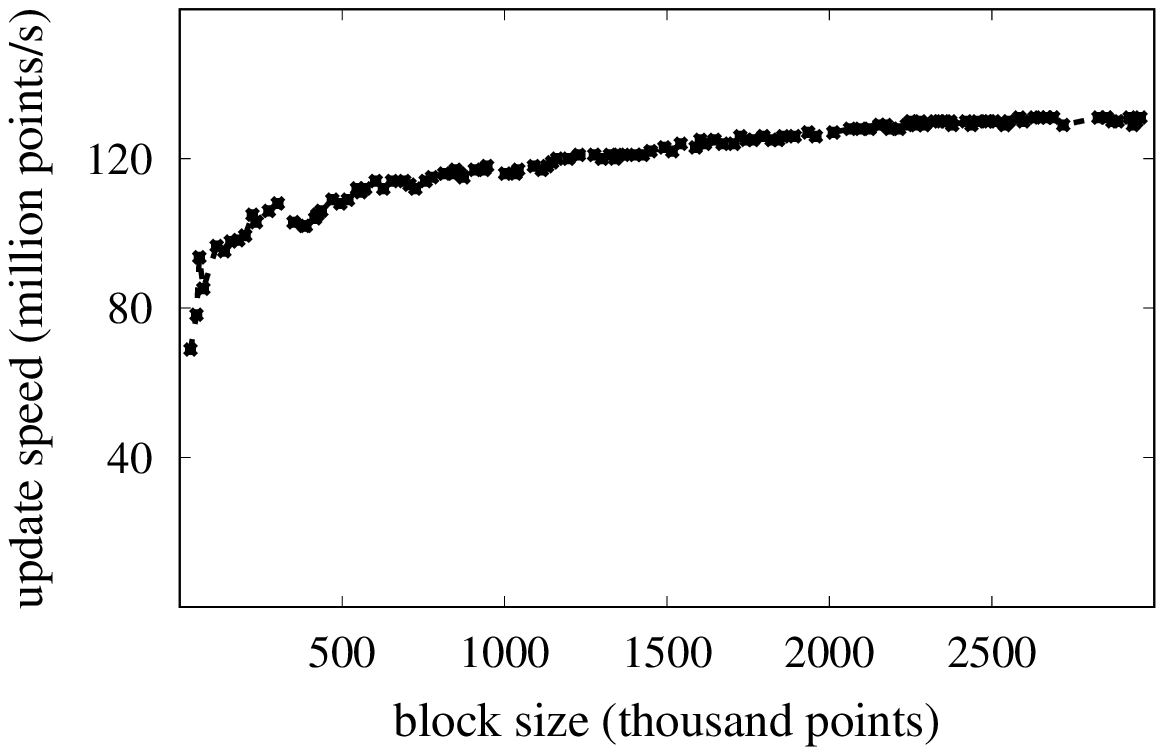}
  \label{fig:cost:calculation_speed_gpu}
}
\hspace{-1em}
\subfigure[CPU]{
  \includegraphics[width=0.5\columnwidth]{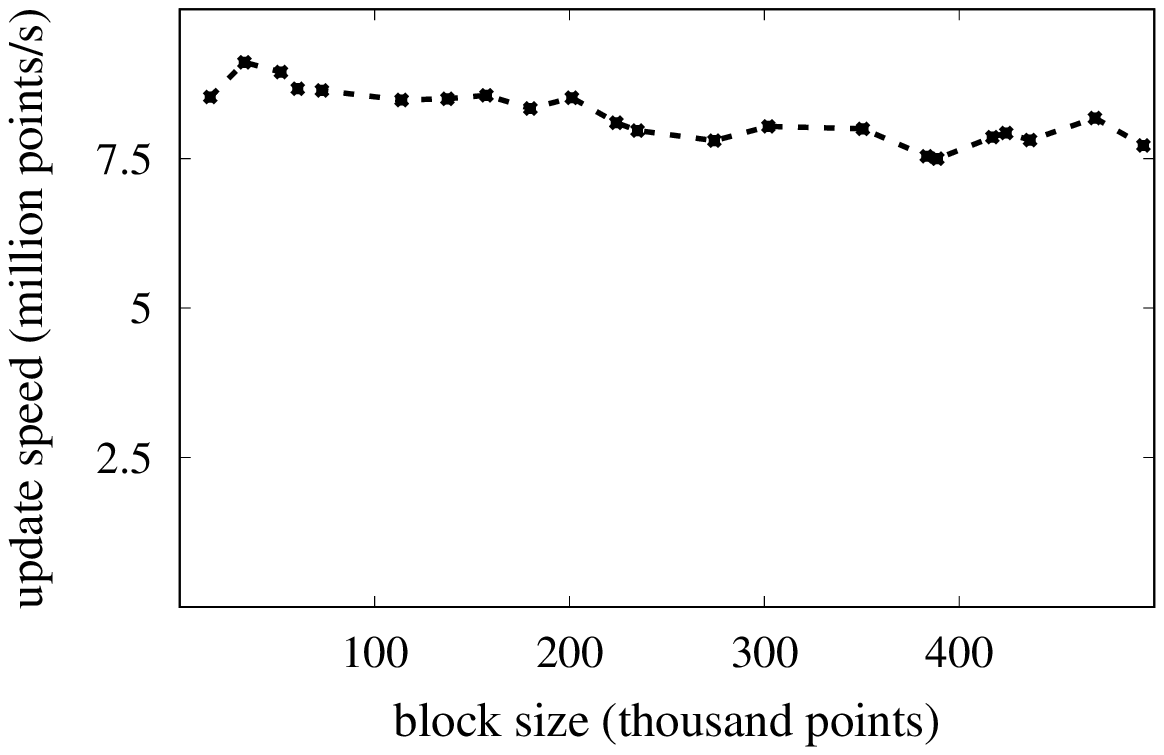}
  \label{fig:cost:calculation_speed_cpu}
}
\end{tabular}
\end{center}
\vspace{-1em}
\caption{Processing speed of GPUs and CPUs on blocks with different sizes}
\label{fig:cost:calculation_speed}
\end{figure}

To indicate the relationship between the block size and the efficiency of the GPU, we launch a GPU kernel with the default configuration to process blocks with different sizes. The details of configuration and hardware can be found in \refsec{exp}. The GPU throughput is reported in \reffig{cost:calculation_speed}(a), where the labels on the x-axis represent the number of elements in a block, and the labels on the y-axis represent the average number of elements processed in every second.

The throughput significantly increases when the block size is relatively small. Afterward, the upward trend becomes gentle as the block size continues to grow. This phenomenon may be due to two main reasons. First, we need to transfer data via the PCI-e bus to the global memory of GPU when launching a GPU kernel. A small block cannot fully utilize the bandwidth of the bus. Second, more data can better utilize all threads and the cache mechanism of the GPU. More throughput details about data transfer and GPU kernel execution can support our opinion and will be shown in \refsubsec{modelgpu}.

\vspace{-0.5em}
\begin{observation}
\label{ob:cpu}
In the context of MF, the computing power of CPU cores is not sensitive to the block size.
\end{observation}
\vspace{-0.5em}

The average number of elements processed by a thread of CPU in every second is shown in \reffig{cost:calculation_speed}(b).
Unlike the GPU, the throughput of a CPU thread always remains stable when the block size varies. This is because the worker threads of CPU are relatively more independent than those of GPU and the computing capability of a CPU thread is not so powerful, compared with the whole GPU device.



\stitle{Nonuniform Matrix Division.} Based on the above two observations, an immediate idea to improve the algorithmic efficiency is to set the block size as large as possible. However, as mentioned in \refsubsec{straightforward_method}, the input matrix should be divided into at least $(\cpunum+\gpunum+1) \times (\cpunum+\gpunum)$ blocks and the block size under this division strategy is still relatively small. For example, we use 16 CPU threads and a GPU in the default configuration of our experiments. Considering the real-world dataset \yh with $1,000,990$ rows and $624,961$ columns, we divide it into at least $18 \times 17$ blocks. Consequently, the number of elements in each block is less than $1$ million, which is still not large enough to saturate the GPU computing power in the light of \reffig{cost:calculation_speed}(a).

To improve the working efficiency of GPUs, we divide the rating matrix into blocks with different sizes. The large ones are assigned to GPUs, and the small ones are assigned to CPUs. Towards this end, there are several issues that we need to address. For example, we need to answer how to set suitable block sizes for both CPU and GPU, and how to divide the rating matrix in practice. We will answer these questions in the following section, and the final matrix division strategy will be given in \refsec{task}.

\stitle{Workload Balance.}
As shown above, we should make the size of blocks assigned to GPUs as large as possible in terms of improving the working efficiency of GPUs. However, an extreme nonuniform division strategy may cause a serious workload imbalance problem, which can remarkably reduce the overall efficiency when combining two hardware resources. To prevent this issue, a scheduling strategy should be considered.
Many efforts have been done to balance workloads for the applications in heterogeneous systems. They can be categorized into three kinds: (1) \textit{dynamic methods}, (2) \textit{static methods by classifier}, and (3) \textit{static methods by cost models}. Here, we briefly review some of representative methods among them and elaborate reasons why they cannot be straightforwardly used in our problem. Then, we propose our own method. For more details of scheduling strategies in heterogeneous systems, interested readers can refer to surveys\cite{pradeep2018review,mittal2015survey}.


\noindent\underline{(1) Dynamic Methods.}
The dynamic methods assign a new task to a computational device according to the performance of devices on previous tasks \cite{rossi2016leveraging,jimenez2009predictive,leis2014morsel,belviranli2013dynamic,boyer2013load,choi2013efficient,wang2013cap,kaleem2014adaptive}. 
For example, \cite{rossi2016leveraging} maintains a double-ended queue. CPU processes the tasks from the front of the queue. Simultaneously, GPU processes the tasks from the reverse direction. The algorithm naturally enables the dominant hardware resource to handle more tasks. 
%
%
%
The baseline algorithm in \refsubsec{straightforward_method} essentially adopts a dynamic strategy. If a CPU core or a GPU finishes updating a block, it is assigned a new block without incurring any conflict. 
The dynamic method performs well in other problems if there is not any rule to assign tasks.

However, this type of method does not work well in our problem due to the independence property of the task assignment.
%
%
As discussed in \refsubsec{straightforward_method}, we first equally divide the matrix into a set of blocks, which is necessary to avoid conflicts. GPUs cannot achieve an ideal performance in this setting according to \refob{small_gpu}. 
In addition to the problem of GPU working efficiency, the dynamic update method in \algnaive may suffer from a poor training result. We give an example to explain this problem.

\begin{figure}[t!]
\begin{center}
\includegraphics[width=0.9\columnwidth]{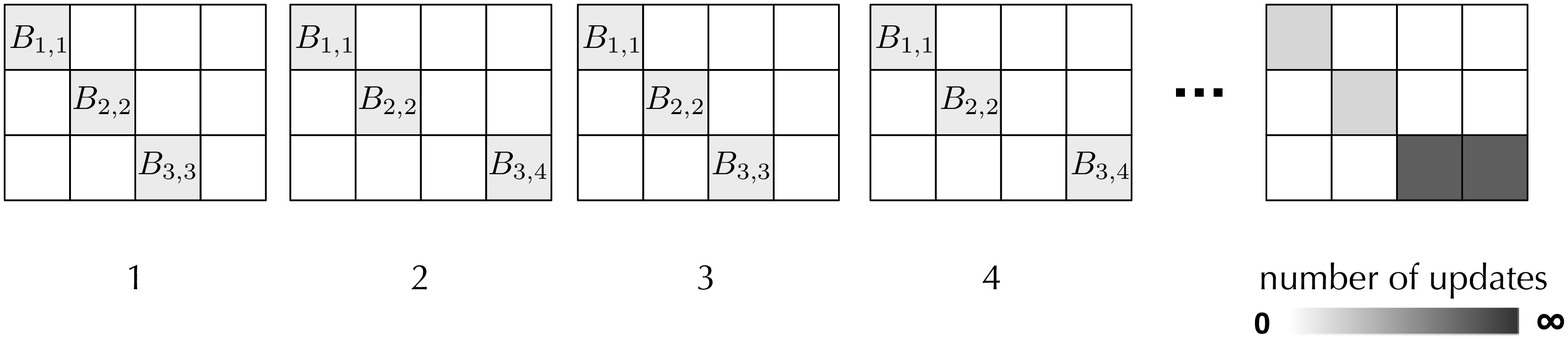}
\vspace{-0.5em}
\caption{A running example of the straightforward algorithm given 2 CPU cores and 1 GPU}
\label{fig:dynamic}
\end{center}
\vspace{-0.5em}
\end{figure}

\vspace{-0.5em}
\begin{example}
\label{ex:poor_training_result}
We consider computing the matrix factorization of a rating matrix $R$ on a machine with two CPU threads $c_1,c_2$ and one GPU $g_1$. Assume that we divide $R$ into $3 \times 4$ matrix blocks, which is shown in \reffig{dynamic}.
In the beginning, two CPU threads $c_1,c_2$ and the GPU $g_1$ get blocks $B_{1,1}$, $B_{2,2}$, and $B_{3,3}$, respectively. Normally, the computing power of a GPU is much stronger than that of a CPU thread. As a result, when $g_1$ has completed its task $B_{3,3}$, $c_1$ and $c_2$ are still working on their blocks $B_{1,1}$ and $B_{2,2}$.
According to the scheduling policy of \algnaive, $g_1$ will apply a new block which is independent of $B_{1,1}$ and $B_{2,2}$ and has the least number of updates. Block $B_{3,4}$ satisfies these conditions. $g_1$ picks $B_{3,4}$ in the second step of \reffig{dynamic}. In the following steps, $g_1$ will continually update the two blocks in the lower right corner since $B_{1,1}$ and $B_{2,2}$ are always occupied by $c_1$ and $c_2$. This phenomenon makes the numbers of updates for different blocks severely unbalanced, which is demonstrated in the last matrix. The number of updates for a block is relatively large if the corresponding color is dark. Compared with the original SGD algorithm which updates matrix elements randomly, the process in this example leads to a weak training result.
\end{example}

\vspace{-0.5em}
\noindent\underline{(2) Static Methods by Classifier.}
Static methods provide scheduling decisions for worker threads of different devices before applications start. This type of method establishes a classifier based on training datasets in the offline phase \cite{grewe2011static,kofler2013automatic,wen2014smart,ghose2016divergence}. Given a new task, the classifier identifies the class of the task and applies a corresponding strategy of task assignment derived from the offline phase.
%


There are several drawbacks if applying the classifier-based methods in our problem. First, it is complicated and time-consuming to generate training data including static code features and optimal partition strategies. Moreover, these methods usually rely on specific frameworks such as OpenCL \cite{stone2010opencl} and Insieme \cite{insieme} to obtain static code features. This makes them not general. Second, they are usually designed for multi-task platforms. Consequently, The partition scheme generated by them is coarse-grained.

\begin{figure}[t!]
\centering
\includegraphics[width=0.9\columnwidth]{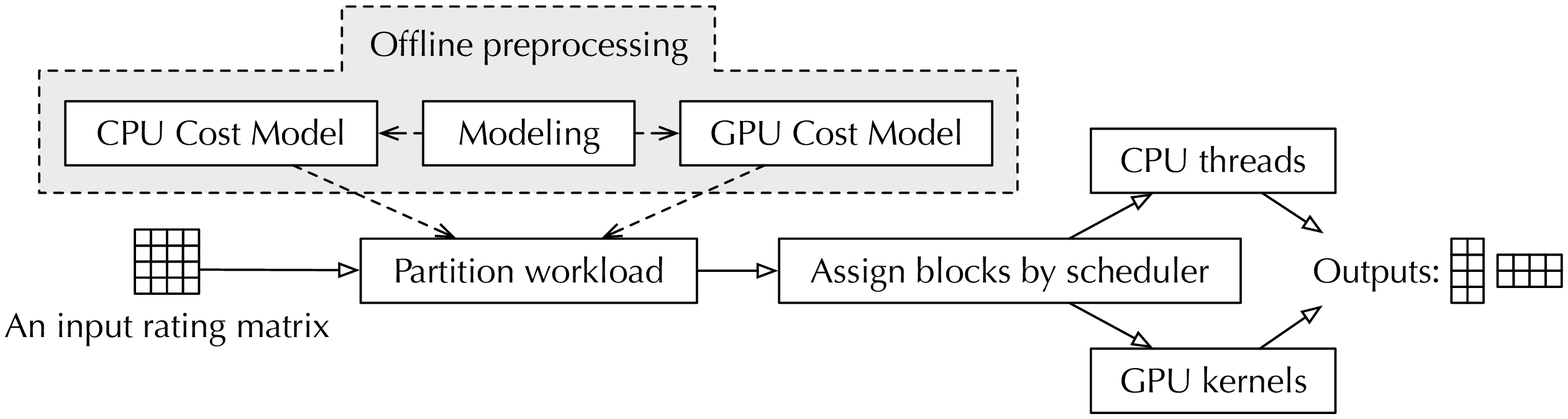}
\vspace{-0.7em}
\caption{Overview of \algopt}
\label{fig:frk}
\end{figure}

\noindent\underline{(3) Static Methods By Cost Model.}
This type of method estimates CPU's and GPU's execution time by establishing a cost model \cite{luk2009qilin,lee2015orchestrating}. A cost model is a function revealing the relationship between the input data size and the corresponding execution time of a specific worker thread.
A representative approach among them is Qilin \cite{luk2009qilin}. It divides a training dataset $N$ into two parts $N_1$ and $N_2$, which are assigned to CPUs and GPUs, respectively. $N_1$ is further divided into $m$ subparts $N_{1,1}$...$N_{1,m}$. Each subpart $N_{1,i}$ is processed by a CPU thread, and the corresponding execution time is recorded. A similar operation is applied to $N_2$ by GPUs. Qilin uses the curve fitting to construct two linear equations as the projections of the execution times for CPUs and GPUs respectively. 

In the context of our problem, a simple linear function in \cite{luk2009qilin} is hard to accurately estimate the execution time of GPUs. Recall \reffig{cost:calculation_speed}(a), it is not a horizontal line. This proves that the execution time does not increase linearly as the number of elements in a block grows.





\stitle{Our Method to Balance Workloads.} We propose a hybrid method for our problem. We first customize a cost model to divide the matrix specialized for the MF problem. We improve the accuracy of the GPU cost model. The details of cost models are given in \refsec{model}. Second, we have a dynamic scheduling mechanism, which allows the dominant resource to use work-stealing mechanism. This alleviates the deviation between the cost model and the practical performance.

\vspace{-0.5em}
\subsection{The Framework}
\label{subsec:frk}
\vspace{-0.2em}

%
In this section, we give an overview of our improved algorithm in \reffig{frk}, which is called \algopt. Our method contains an offline preprocessing phase and an online processing phase. The offline phase (the gray area in \reffig{frk}) derives a cost model which estimates the hardware performance. This step can be performed only once on a machine, and the corresponding parameters are stored to support the query of any input rating matrix in the online phase.

In the online phase, a sparse rating matrix $R$ is given. \algopt first divides the matrix into two parts, denoted as $R_1$ and $R_2$, based on cost models of CPUs and GPUs from the offline phase. Then, $R_1$ and $R_2$ are further divided into several blocks. Finally, the scheduler assigns blocks to worker threads. The calculation process continues until the number of iterations reaches the predefined value. During most of the period when \algopt runs, CPU threads are only allowed to process blocks in $R_1$, and GPUs are only allowed to process blocks in $R_2$. Similar to \algnaive, the block assignment avoids conflicts in the same row or column. We also have a dynamic scheduling strategy to balance workloads in practice, which is not reflected in \reffig{frk}. The details will be shown in \refsec{task}. A formal pseudocode of the framework \algopt is reported in \refalg{hmf}, which is self-explanatory.

\begin{algorithm}[t!]
\caption{\algopt}
\label{alg:hmf}
\KwIn{$R_{m\times n}, k, \lambda_{P}, \lambda_{Q}, \gamma, t, \cpunum, \gpunum$}
\KwOut{$P_{m \times k}, Q_{k \times n}$}

\tcp{Offline Phase}
generate cost models of both CPUs and GPUs\;
\tcp{Online Query Processing}
partition $R_{m\times n}$ according to the cost models\;
preprocess data\;
$\kw{init\_scheduler}(R_{m\times n}, \cpunum, \gpunum)$\;
scheduler assigns blocks to CPU threads and GPUs\;
\Return $P_{m \times k}, Q_{k \times n}$\;
\end{algorithm}


\section{Our Cost Model}
\label{sec:model}
\vspace{-0.5em}
Given an input matrix $R$, let $\alpha$ and $1-\alpha$ be the proportion of the workload assigned to GPUs and CPUs, respectively, where $0 \le \alpha \le 1$. We use $T_g(\alpha \cdot R)$ and $T_c((1-\alpha) \cdot R)$ to denote the time spent on updating elements in $\alpha \cdot R$ and $(1-\alpha) \cdot R$ by a GPU and a CPU thread, respectively. When the context is clear, $T_g(\alpha)$ and $T_c(1-\alpha)$ are used for short. The total running time $T$ of our algorithm is represented below.

\begin{equation}
\label{eq:optimal_speed_up}
T = \max(\frac{T_g(\alpha)}{\gpunum},\frac{T_c(1-\alpha)}{\cpunum})
\end{equation}
\vspace{-1em}

Note that both $T_g(\alpha)$ and $T_c(1-\alpha)$ are monotonic. Based on the computed cost functions, the total running time is minimized when the load between resources keeps balancing. We set $\alpha$ using the following equation.

\begin{equation}
\label{eq:optimal_load}
\alpha = argmin |\frac{T_g(\alpha)}{\gpunum}-\frac{T_c(1-\alpha)}{\cpunum}|
\end{equation}
\vspace{-1em}

Based on discussion above, our aim is to derive cost functions $f_g(\alpha) \simeq T_g(\alpha)$ and $f_c(\alpha) \simeq T_c(\alpha)$ for a GPU and a CPU thread, respectively, where $f_g(\alpha)$ (resp. $f_c(\alpha)$) denotes the estimation of $T_g(\alpha)$ (resp. $T_c(\alpha)$). As discussed earlier, a straightforward method to establish a cost model is to follow the works \cite{luk2009qilin,lee2015orchestrating} which think that execution time is linearly related to the size of the input matrix. However, based on \refob{small_gpu}, we find that the processing speed of GPUs increases when the block size increases in our problem. This makes linear regression methods for the GPU cost model inaccurate. Moreover, the computation in execution kernels of GPUs and the data transfer are not completely serial due to CUDA stream mechanism. The total running time of GPU is not a simple sum of the kernel execution and the data transfer. This observation makes us reconsider how execution kernel and data transfer exactly influence the total running time of GPUs, which is not discussed in previous cost models.

In the rest of this section, we introduce the strategy to prepare the training data in \refsubsec{prepare} and propose our cost model of GPUs in \refsubsec{modelgpu}. For the cost model of CPUs, we use a linear function to estimate the performance similar to \cite{luk2009qilin}. A formal pseudocode to compute cost models is given in \refalg{cost_model}. We explain each step as follows.

\begin{algorithm}[t!]
  \caption{Cost Estimation\xspace}
  \label{alg:cost_model}
  \KwIn{$R_{m\times n}$}
  \KwOut{$f_c$ and $f_g$}
  \tcp{S is an array of segments}
  \tcp{$P_{c}$ and $P_{g}$ are arrays of structures}
  \tcp{Data preprocessing phase}
  $S = \kw{sample\_dataset}(R_{m\times n})$\;
  \tcp{Training cost models}
  $P_{c} = \kw{test\_cpu\_kernel}(S)$\;
  $f_c = \kw{cpu\_model\_fitting}(P_{c})$\;
  $f_g^{transfer} = \kw{test\_transfer\_speed}()$\;
  $P_{g} = \kw{test\_gpu\_kernel}(S)$\;
  $f_g^{execute} = \kw{gpu\_model\_fitting}(P_{g})$\;
  $f_g = \kw{combine}(f_g^{transfer}, f_g^{execute})$\;
\end{algorithm}

\vspace{-0.5em}
\subsection{Data Preparation and Training for CPUs}
\label{subsec:prepare}
\vspace{-0.2em}

To derive the training data, we shuffle the input dataset to avoid uneven data distribution. After the data is shuffled, we equally divide input dataset into $N$ disjoint parts $S_1, S_2, S_3...S_N$, stored in array $S$ at line 1 of \refalg{cost_model}. Then, CPU execution kernel configured with a single thread is launched to compute on datasets generated in the last step at line 2. Instead of computing on $S_1, S_2, S_3...S_N$ respectively in \cite{luk2009qilin}, CPU execution kernel computes on $S_1, S_1+S_2, S_1+S_2+S_3...S_1+S_2+S_3+...+S_N$ respectively, and the corresponding execution time is recorded. After this, we get an array $P_{c}$ including data size and corresponding execution time. As a training data set, this array is used to curve fitting for CPUs. Our adaptation generates a wider range of training data which can better reflect the relationship between data size and execution time. To eliminate noise, the execution time in the training data is derived from the average of multiple tests.


\subsection{Estimating Working Efficiency of GPUs}
\label{subsec:modelgpu}

Given a task, the total processing time by a GPU is spent on two parts: (1) \textit{data transfer between the CPU and the GPU via the PCI-e bus}, and (2) \textit{GPU execution kernels}.


\stitle{Data Transfer.} Data transfer has two directions --- from CPU to GPU (sometimes called Host to Device) and from GPU to CPU. We denote the times spent on them by $f_g^{c \Rightarrow g}$ and $f_g^{g \Rightarrow c}$, respectively. We only discuss the process to model the data transfer from CPU to GPU, and the model for the other direction is similar.

\begin{figure}[t!]
\begin{center}
\begin{tabular}[t]{c}
\hspace{-1em}
\subfigure[CPU to GPU]{
  \includegraphics[width=0.5\columnwidth]{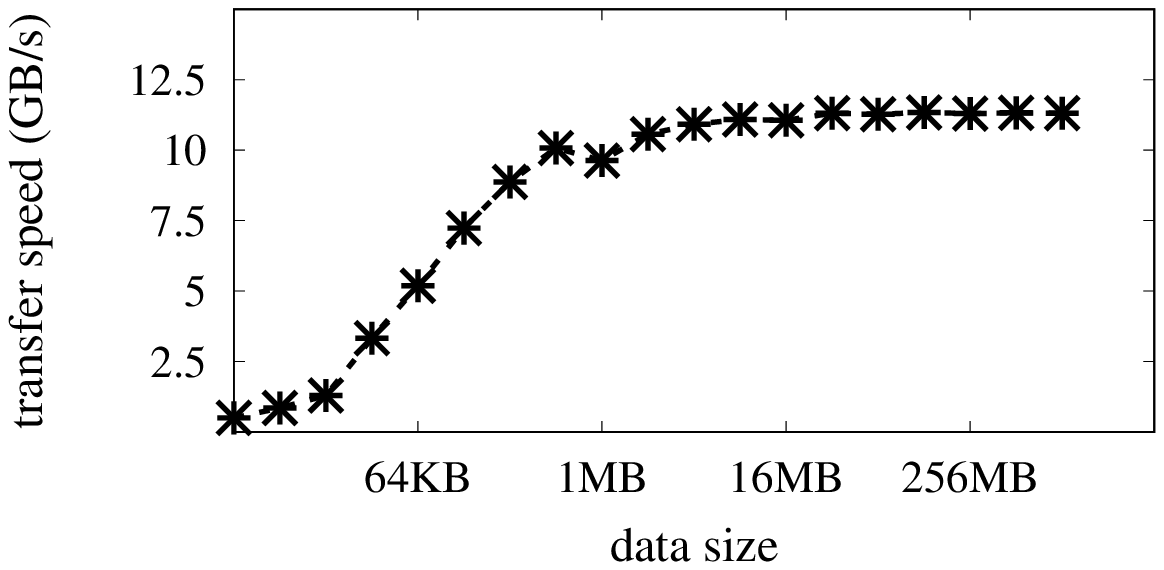}
}
\hspace{-1em}
\subfigure[GPU to CPU]{
  \includegraphics[width=0.5\columnwidth]{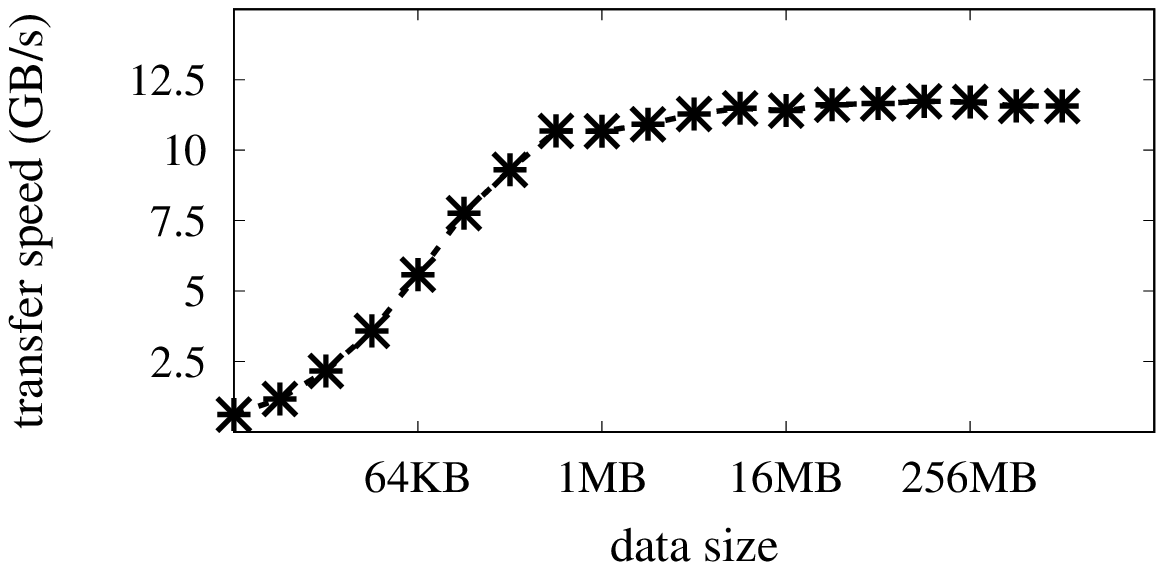}
}

\end{tabular}
\end{center}
\vspace{-1em}
\caption{Transfer speed varies with block size}
\label{fig:cost:transfer_speed}
\end{figure}

\reffig{cost:transfer_speed} reveals the transfer speed for data with different sizes. The transfer speed grows very fast in the beginning. After the data size is larger than a threshold, the transfer speed remains stable. Based on this phenomenon, we use a function to fit the curve when the dataset is not very large, then use the linear regression to model the rest. A formal model is expressed as follows. $|R|$ denotes the size of data transferred from CPU to GPU, and $\tau$ denotes the threshold.

\vspace{-0.5em}
$$f_g^{c \Rightarrow g}=\left\{
\begin{aligned}
\frac{|R|}{a \cdot \sqrt{log|R|}+b} & & \text{if } |R| \le \tau; \\
a \cdot |R|+b & & \text{otherwise}.
\end{aligned}
\right.
$$
\vspace{-0.5em}

The rationale for $|R| \le \tau$ is that the data transfer time can be represented by the quotient of the data size and the transfer speed. According to \reffig{cost:transfer_speed}, we use the function $a \cdot \sqrt{log|R|}+b$ to model the curve of the first stage. 
We select this function since the trend performs like an inverse function of the parabola. Note that the label distribution on the x-axis is not linear. This is because the logarithmic scale can make the trend clearly presented even though the data size is small. Obviously, this phenomenon shows that we cannot fully utilize the bandwidth of the PCI bus if the data is not large enough, which supports \refob{small_gpu}. 
Then, we determine the threshold $\tau$ by the extent to transfer speed variation. Empirically, when the variation of the transfer speed is less than 2\% in a time unit, we consider that the transfer speed has been stable. 
Finally, we fit the curve as a linear function for the second stage when $|R| > \tau$. The curve fitting can be done by using the least squares method.

\begin{figure}[t!]
\begin{center}
\includegraphics[width=0.5\columnwidth]{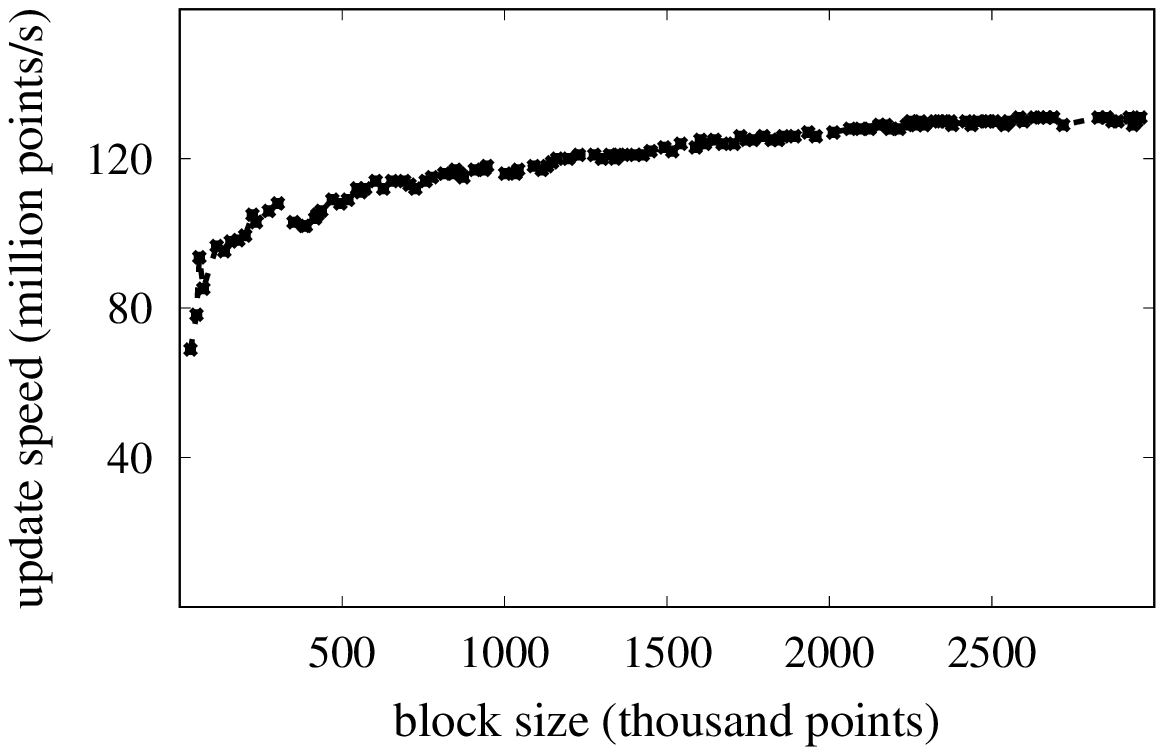}
\end{center}
\vspace{-1em}
\caption{Kernel execution time by varying data size}
\label{fig:kernel}
\end{figure}


\stitle{GPU Execution Kernel.} We design the cost model of the GPU execution kernel.
Similar to \reffig{cost:transfer_speed}, the throughput of updating remains stable after the block size reaches a threshold, which means that the computing power of the GPU is saturated. To fit the curves, we use a logarithmic function when the dataset is not very large, and then use the linear regression to model the rest. The growth trend of the logarithmic function can be slower than the power function (e.g., $\sqrt{x}$), which is more consistent with the trend in \reffig{kernel}. This is why we choose it to model the curve of the first stage. A formal model is expressed as follows, where $f_g^{kernel}$ denotes the time spent on $R$ by the GPU execution kernel. The function $a \cdot log|R|+b$ represents the processing speed of the GPU execution kernel.

\vspace{-1em}
$$f_g^{kernel}=\left\{
\begin{aligned}
\frac{|R|}{a \cdot log|R|+b} & & \text{if } |R| \le \tau; \\
a \cdot |R|+b & & \text{Otherwise}. 
\end{aligned}
\right.
$$
\vspace{-0.5em}


\stitle{Overall GPU Cost Model.} We cannot simply sum the time of the kernel execution and the data transfer as our estimation for the overall execution time of the GPU, since these two parts are not absolutely serial. Specifically, to improve the overall working efficiency of GPUs, we adopt a widely used optimization based on the CUDA stream mechanism. A CUDA stream contains a list of GPU commands executed in serial, and commands in different streams are executed in parallel if hardware resources permit. At the same time, commands in different streams can be synchronized. This mechanism allows us to perform data transfer and kernel execution in parallel without breaking correctness. We explain this idea in the following example.

\vspace{-1.5em}
\begin{figure}[ht]
\begin{center}
\includegraphics[width=0.95\columnwidth]{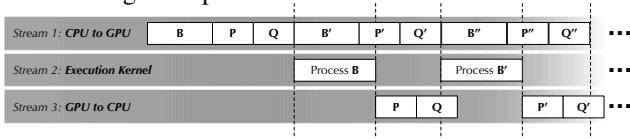}
\vspace{-0.5em}
\caption{Data transfer optimization}
\label{fig:overlap_technique}
\end{center}
\end{figure}
\vspace{-2em}

\begin{example}
As shown in \reffig{overlap_technique}, we use three streams to manage the data transfer from CPU to GPU, the kernel execution, and the data transfer from GPU to CPU, respectively. Assume that a block $B$ is assigned to the GPU. Stream 1 transfers $B$ and corresponding $P,Q$ to the global memory of a GPU. Then, the GPU kernel scans the block $B$ and updates $P,Q$. Simultaneously, stream 1 continuously transfers the next block $B'$ assigned to the GPU and corresponding $P',Q'$. When stream 2 finishes $B$, stream 3 transfers the updated $P,Q$ back to CPU.
\end{example}


From this example, the overall time for GPU $f_g$ can be roughly decided by the maximum time spent among these three streams because it covers the time of the other two parts. Note that although the first and the last schemes cannot be overlapped by the maximal stream in the figure, the cost can be ignored when the number of transferred and computed blocks is very large. Note that $f_g^{g \Rightarrow c}$ is always smaller than $f_g^{c \Rightarrow g}$ since we do not need to transfer blocks back to CPU. Therefore, we define the overall cost model of a GPU as follows.

\vspace{-0.5em}
\begin{equation}
\label{eq:finalgpu}
f_g=\max(f_g^{c \Rightarrow g}, f_g^{kernel})
\end{equation}
\vspace{-1em}

The overall cost model of a GPU depends on the maximum between data transfer time from CPU and GPU and execution time of the GPU kernel.





\section{Workload Balance in Practice}
\label{sec:task}
\vspace{-0.2em}
\subsection{Dynamic Scheduling}
\vspace{-0.2em}

Even though we have proposed a tailored GPU cost model for MF, the estimation may be still hard to exactly reflect the computing power of devices given a different dataset. The workloads of CPU and GPU may be unbalanced if we assign blocks simply according to the cost model. To remedy this issue, we adopt a dynamic scheduling strategy when one device finishes its tasks. Specifically, assume that the GPU has finished its tasks. Instead of waiting for the tasks being processed by CPUs, we allow GPUs to pick some blocks originally assigned to CPUs. We call it \textit{static phase} when the GPU and the CPU only process the originally assigned tasks and call it \textit{dynamic phase} when one of them finishes its own tasks and is involved in processing tasks of the other. In \textit{static phase}, every GPU is assigned to blocks in specific rows so that it can update one segment of one result matrix all the time and avoid the transfer of this segment.

\vspace{-0.2em}
\subsection{Putting Things Together}
\vspace{-0.2em}

\begin{figure}[t!]
\begin{center}
\includegraphics[width=0.99\columnwidth]{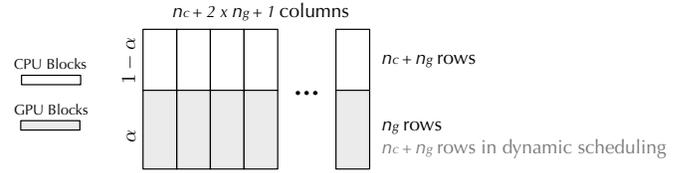}
\caption{The final division strategy}
\label{fig:final_division}
\end{center}
\vspace{-0.5em}
\end{figure}

We explain our final strategy for the matrix division in this section. An example is shown in \reffig{final_division}.

\stitle{Number of Columns.} Based on the cost model proposed in \refsec{model}, we first partition the matrix into two sub-matrices $R_c$ and $R_g$ for CPUs and GPUs, respectively. They are marked by white and gray in \reffig{final_division}. In the light of \refrul{rule1}, we further divide the matrix into $\cpunum+2\times\gpunum+1$ columns. This setting guarantees two things. The first thing is that GPUs can always know not only the current block but also the next block. This enables the overlap between computation and data transfer in \reffig{overlap_technique}. The second thing is that there always exists a spare column when a GPU kernel or a CPU thread finishes processing its block.

\stitle{Number of Rows for CPUs.} As shown in \refrul{rule1}, we can divide the input matrix into $\cpunum+\gpunum$ rows, where there are $\cpunum$ (resp. $\gpunum$) rows in $R_c$ (resp. $R_g$). However, this division strategy causes a problem when the dynamic scheduling is activated. Specifically, assume that GPUs have finished their own tasks and are involved in processing blocks of CPUs. Currently, we have totally $\cpunum+\gpunum$ (CPU and GPU) threads working on $(\cpunum+2\times\gpunum+1) \times \cpunum$ blocks. This would break \refrul{rule1} and cannot fully exploit all worker threads. To support the assistance from GPUs, we set the number of rows of CPUs as $\cpunum+\gpunum$ based on \refrul{rule1}. The setting would not affect the CPU efficiency since the computing power of CPUs is not sensitive to the block size as shown in \refob{cpu}.

\stitle{Number of Rows for GPUs.} Similarly, If CPUs first finish their tasks and apply for blocks in $R_g$, the number of rows in $R_g$ should be at least $\cpunum+\gpunum$. However, compared with the row number $\gpunum$ for GPUs, the row number $\cpunum+\gpunum$ leads to a smaller block size, which cannot saturate the computing power of GPUs according to \refob{small_gpu}. Different from the case of CPUs, the division strategy for GPUs needs to satisfy that the block size is large enough in the beginning, while the number of rows is large enough to avoid conflicts if CPUs join. To achieve this target, we divide $R_g$ into $\gpunum$ rows. For each row $R_g^i$ where $1 \le i \le \gpunum$, we further divide $R_g^i$ into $\lceil \frac{\gpunum+\cpunum}{\gpunum} \rceil$ sub-rows. As a result, relatively large blocks with sizes $\frac{R_g}{(\cpunum+2\times\gpunum+1) \times \gpunum}$ are assigned to GPUs in static phase, and blocks with sizes $\frac{R_g}{(\cpunum+2\times\gpunum+1) \times \gpunum \times \lceil \frac{\gpunum+\cpunum}{\gpunum} \rceil}$ are assigned to GPUs and CPUs in dynamic phase.


\begin{example}
We give an example to explain the division strategy for $R_g$. Assume that we have 2 GPUs and 4 CPU threads, i.e., $\cpunum = 4, \gpunum = 2$. We divide $R_g$ into $2$ rows, and each row is further divide into $3$ sub-rows. In static phase, we assign a block with size $\frac{R_g}{9 \times 2}$ to a GPU, and in dynamic phase, we assign a block with size $\frac{R_g}{9 \times 6}$ to a GPU or a CPU thread. On the other hand, $R_c$ is divided into $9$ columns and $6$ rows. This division for $R_c$ would not change in the algorithm.
\end{example}


\section{Experiments}
\label{sec:exp}
We conduct extensive experiments to show the efficiency and the effectiveness of our proposed algorithms. Algorithms appearing in our experiments are summarized as follows.

\vspace{-0.2em}
\begin{itemize}
\item \algcpu: Only CPU works. We uniformly divide the matrix and use the strategy in \cite{zhuang2013fast} to assign blocks. More details can be found in \refsubsec{cpu_algorithm}. We use AVX and OpenMP for acceleration.
\item \alggpu: Only GPU works. We vary the number of rows and columns for the matrix division and adopt the best one. The "-O3" optimization flag is supported. 
\item \algnaive: CPU and GPU work in parallel. The algorithm is introduced in \refsubsec{straightforward_method}. AVX, OpenMP, and "-O3" optimization flag are supported.
\item \algopt: CPU and GPU work in parallel. Nonuniform matrix division and dynamic strategy are used. Our cost model decides the size of blocks assigned to two hardware resources. AVX, OpenMP, and "-O3" optimization flag are supported.
\end{itemize}
\vspace{-0.2em}

Stochastic gradient methods and the parameter $k$ used in \cite{zhuang2013fast} and \cite{xie2017cumf_sgd} are different. To correctly combine two methods on Heterogeneous CPU-GPU Systems, we embed the core part of \textbf{LIBMF}\footnote{\url{https://github.com/cjlin1/libmf}} and \textbf{CuMF\_SGD}\footnote{\url{https://github.com/cuMF/cumf_sgd}} into our code and make minor modifications to make the stochastic gradient methods they use consistent. For stochastic gradient method, We choose to use the more concise one in \cite{zhuang2013fast}. For the value of parameter $k$, we choose the larger one in \cite{xie2017cumf_sgd} because a large $k$ value can lead to a better training result.

\stitle{Datasets and Parameter Setting.}
We evaluate algorithms in four real-world datasets --- \textit{\ml}\footnote{\url{http://grouplens.org/datasets/movielens/}},
\textit{\nf}\footnote{\url{https://www.kaggle.com/netflix-inc/netflix-prize-data}},
\textit{\ro}\footnote{\url{https://webscope.sandbox.yahoo.com/catalog.php?datatype=r}}, and 
\textit{\yh}\footnote{\url{https://webscope.sandbox.yahoo.com/catalog.php?datatype=c}}. Statistics of the datasets are presented in \reftab{dataset}. For reproducibility, we consider the original training/test sets in our experiments. More details about each dataset can be found on the corresponding website. We set the parameters following~\cite{chin2015learning}, which are also listed in \reftab{dataset}.

\begin{table}[t]
\scriptsize
\centering
\caption{Network statistics and parameter settings}
\vspace{-0.5em}
\label{tab:dataset}
\begin{tabular}{c|c|c|c|c}
\hline \textbf{Datasets}     & \ml & \nf      & \ro             & \yh \\\hline
\hline \textbf{$m$}          & 71567     & 2649429      & 1948883        & 1000990     \\
\hline \textbf{$n$}          & 65133     & 17770        & 1101750        & 624961      \\
\hline \textbf{$\#Training$} & 9301274   & 99072112     & 104215016      & 252800275   \\
\hline \textbf{$\#Test$}     & 698780    & 1408395      & 11364422       & 4003960     \\
\hline \textbf{$k$}          & 128       & 128          & 128            & 128         \\
\hline \textbf{$\lambda_{P}$}& 0.05      & 0.05         & 1              & 1           \\
\hline \textbf{$\lambda_{Q}$}& 0.05      & 0.05         & 1              & 1           \\
\hline \textbf{$\gamma$}     & 0.005     & 0.005        & 0.005          & 0.01        \\
\hline
\end{tabular}
\vspace{0.5em}                                                                      
\end{table}

\stitle{Experimental Environment.}
We use a machine with Intel Xeon E5-2687W v3 3.10GHz processors and a Quadro P4000 GPU with 8GB global memory. The number of available cores is 20. The system interface of the GPU is PCI Express 3.0$\times$16. The total bandwidth is 32GB/s. By default, we use 16 CPU threads and 128 GPU parallel workers. Here, we follow the definition of GPU parallel workers in \cite{xie2017cumf_sgd}, which means the number of elements computed simultaneously in the GPU kernel. All datasets can fit in memory in our experiments.

\stitle{Organization.}
\refsubsec{exp:overall} shows the adaptiveness of our algorithms by varying the computing resources. \refsubsec{exp:training} shows the effectiveness of our algorithm compared with the state-or-the-art competitor. \refsubsec{exp:opt} and \refsubsec{exp:balance} evaluate our optimization techniques including matrix division strategy and workload balance.

\vspace{-0.5em}
\subsection{Overall Efficiency}
\label{subsec:exp:overall}
\vspace{-0.2em}

We evaluate the overall efficiency of our final algorithm \algopt with \algcpu and \alggpu as comparisons. We use Root Mean Square Error (RMSE) \footnote{\url{https://en.wikipedia.org/wiki/Root-mean-square_deviation}} as a metric for the loss, which is widely used in many recommender systems. For each dataset, we terminate all algorithms and record the corresponding running time when the RMSE reaches a predefined value. Given that we use a different stochastic gradient method from \cite{xie2017cumf_sgd} and a different \textit{k} value from \cite{zhuang2013fast}, the predefined loss values they used are not available. For fair comparison, we select these values that can be reached by all methods including \algnaive which suffers from a weak training quality. The predefined loss values are $0.66$, $0.82$, $20$, and $19$ for \ml, \nf, \ro, and \yh, respectively. The comparisons between \algnaive and \algopt will be shown in \refsubsec{exp:opt}.

\subsubsection{Varying GPU parallel workers}

\begin{figure}[t!]
\begin{center}
\begin{tabular}[t]{c}
\subfigure{
  \includegraphics[width=0.9\columnwidth]{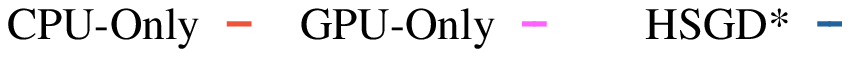}
}
\vspace{-0.5em}
\\
\setcounter{subfigure}{0}
\hspace{-1em}
\subfigure[\ml]{
  \includegraphics[width=0.5\columnwidth]{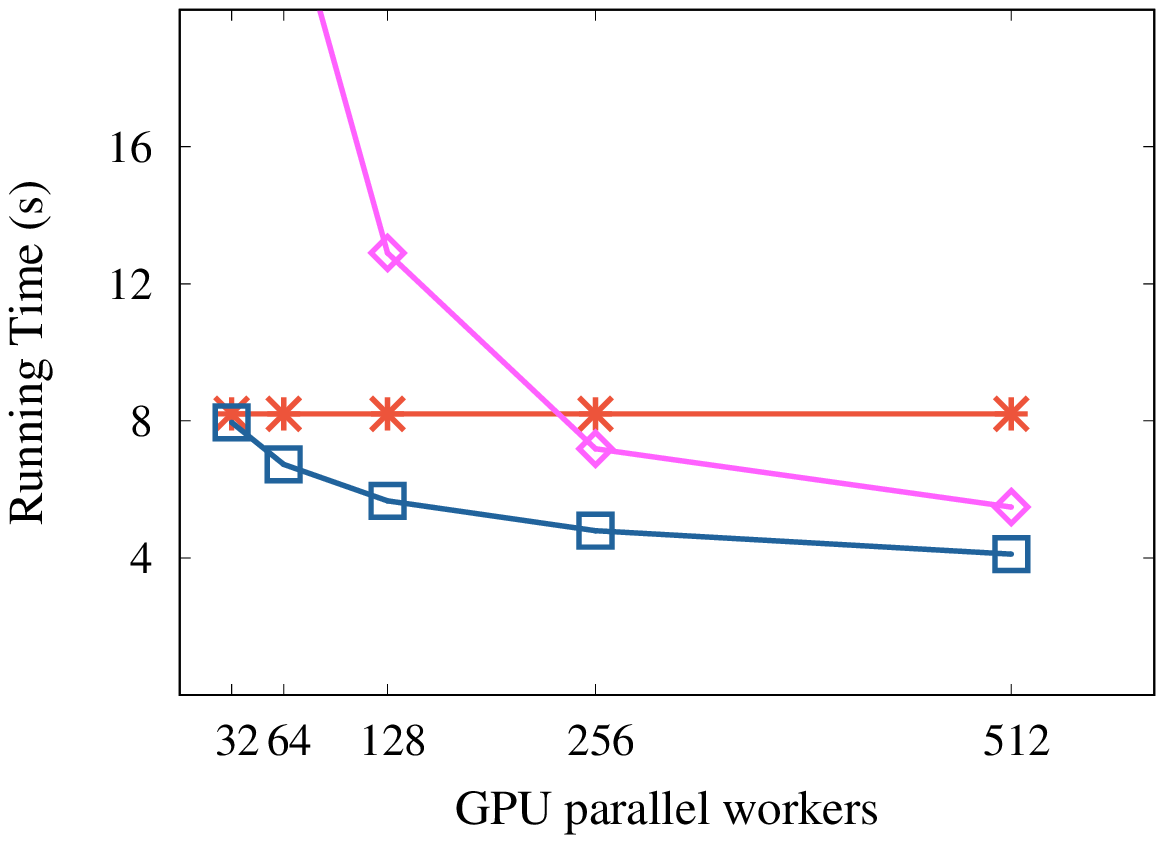}
}
\hspace{-1em}
\subfigure[\nf]{
  \includegraphics[width=0.5\columnwidth]{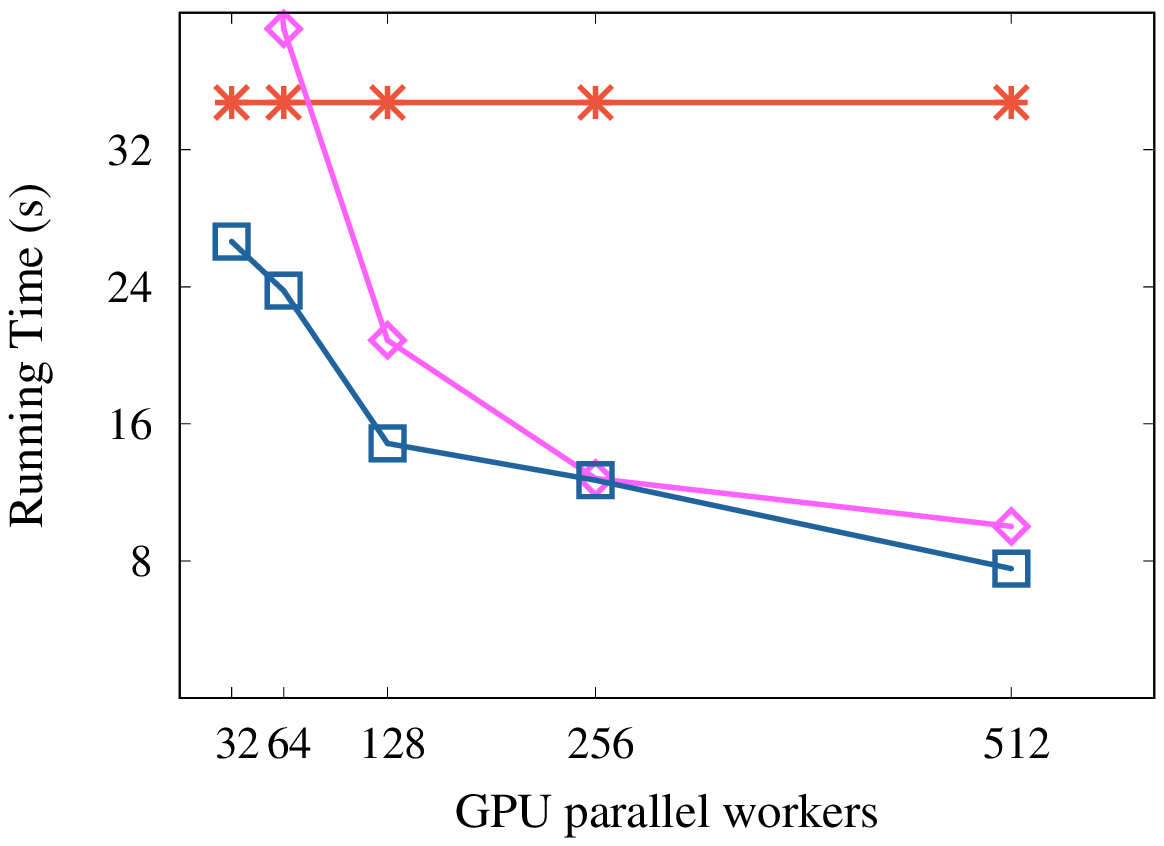}
}
\\
\hspace{-1em}
\subfigure[\ro]{
  \includegraphics[width=0.5\columnwidth]{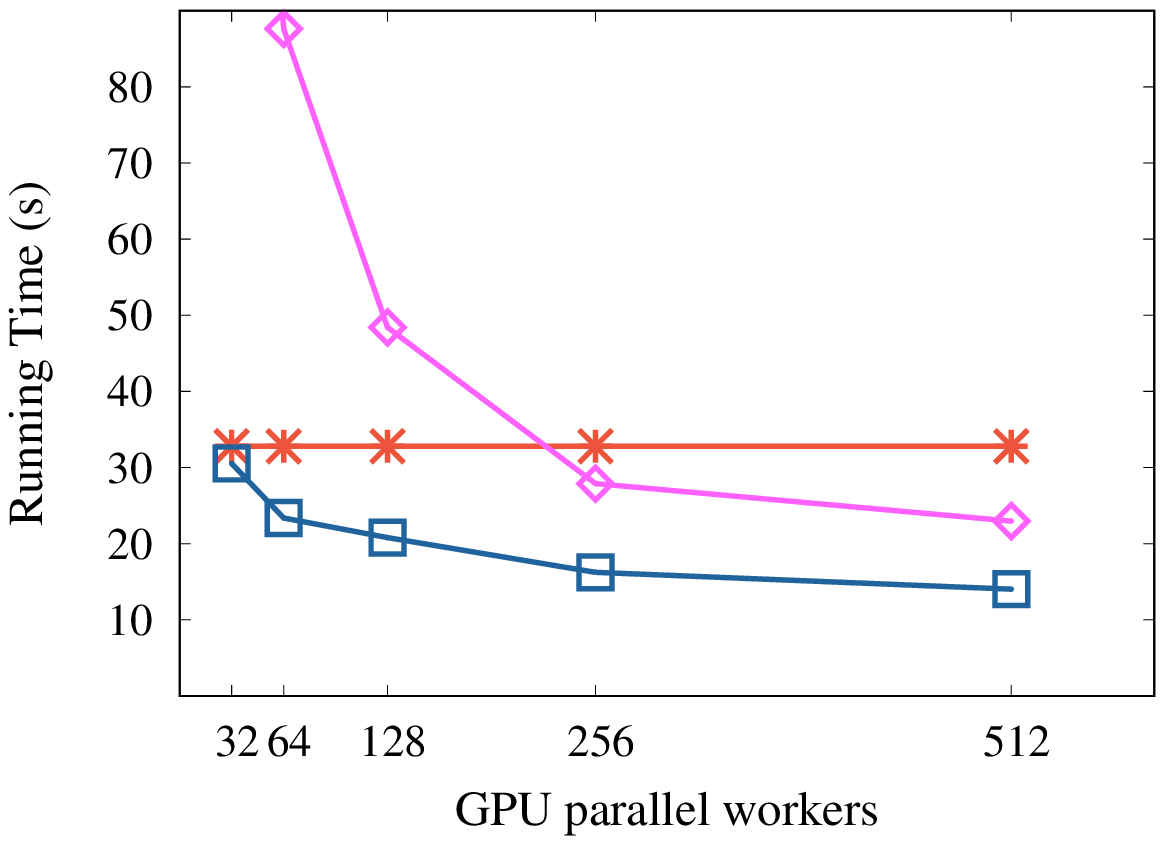}
}
\hspace{-1em}
\subfigure[\yh]{
  \includegraphics[width=0.5\columnwidth]{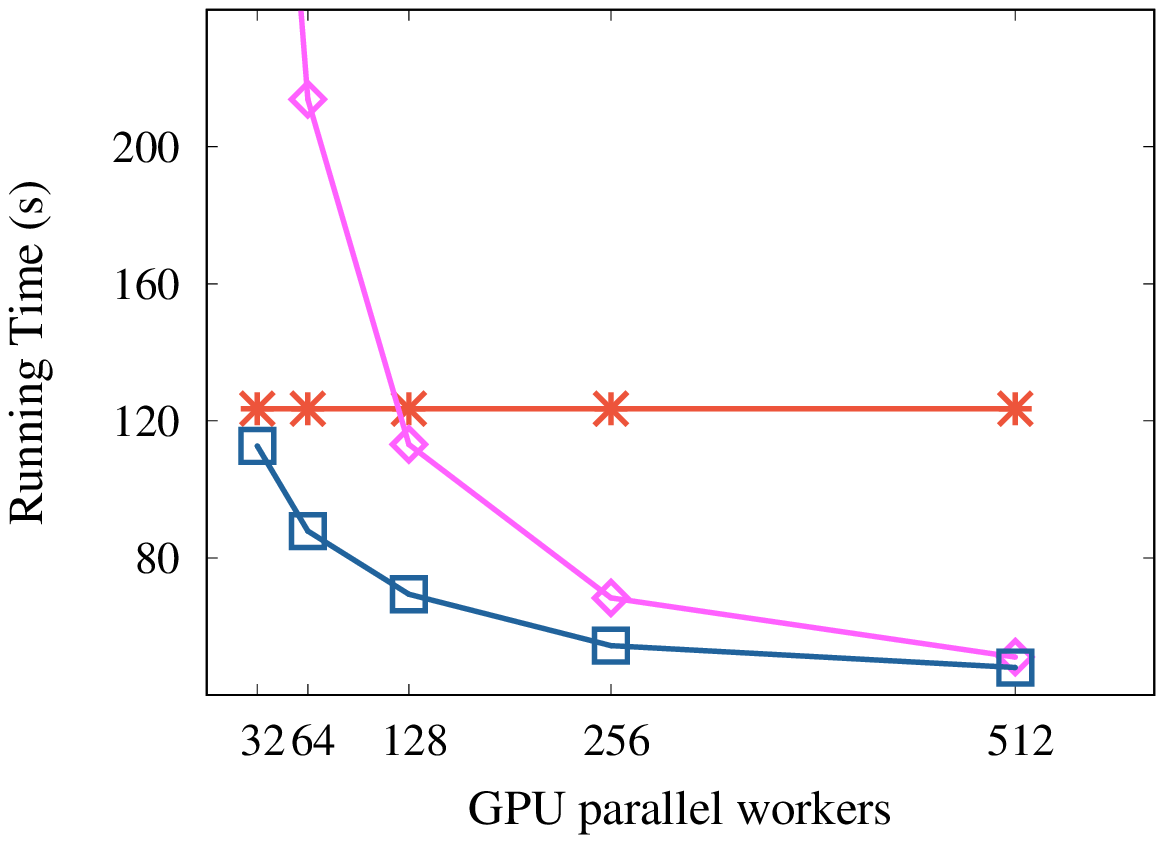}
}

\end{tabular}
\end{center}
\vspace{-1em}
\caption{Varying GPU Threads}
\label{fig:exp:gpu}
\end{figure}

In this experiment, we evaluate the adaptiveness of our algorithm by varying the GPU parallel workers from 32 to 512. The running times of algorithms for different GPU parallel workers are reported in \reffig{exp:gpu} for four datasets. Note that the CPU thread number is fixed to the default value $16$.

As a reference, the running time of \algcpu is stable on all settings. Initially, the \alggpu is slower than \algcpu. When we use more GPU threads, the running time of \alggpu decreases and overtakes that of \algcpu. The running time of \algopt is the smallest among all algorithms. For example, in \ro, given $32$ GPU worker threads, \algopt takes $30$ seconds while \algcpu and \alggpu take $33$ seconds and $170$ seconds respectively. When the thread number increases to 512, \algopt takes $14$ seconds while \alggpu takes $23$ seconds. The decrease of the time of \algopt also shows that our algorithm is adaptive to different GPU settings and can fully utilize the increasing computing power of GPUs.

\subsubsection{Varying CPU Thread Number}

\begin{figure}[t!]
\begin{center}
\begin{tabular}[t]{c}
\subfigure{
  \includegraphics[width=0.9\columnwidth]{exp_dw/key.eps}
}
\vspace{-0.5em}
\\
\setcounter{subfigure}{0}
\hspace{-1em}
\subfigure[\ml]{
  \includegraphics[width=0.5\columnwidth]{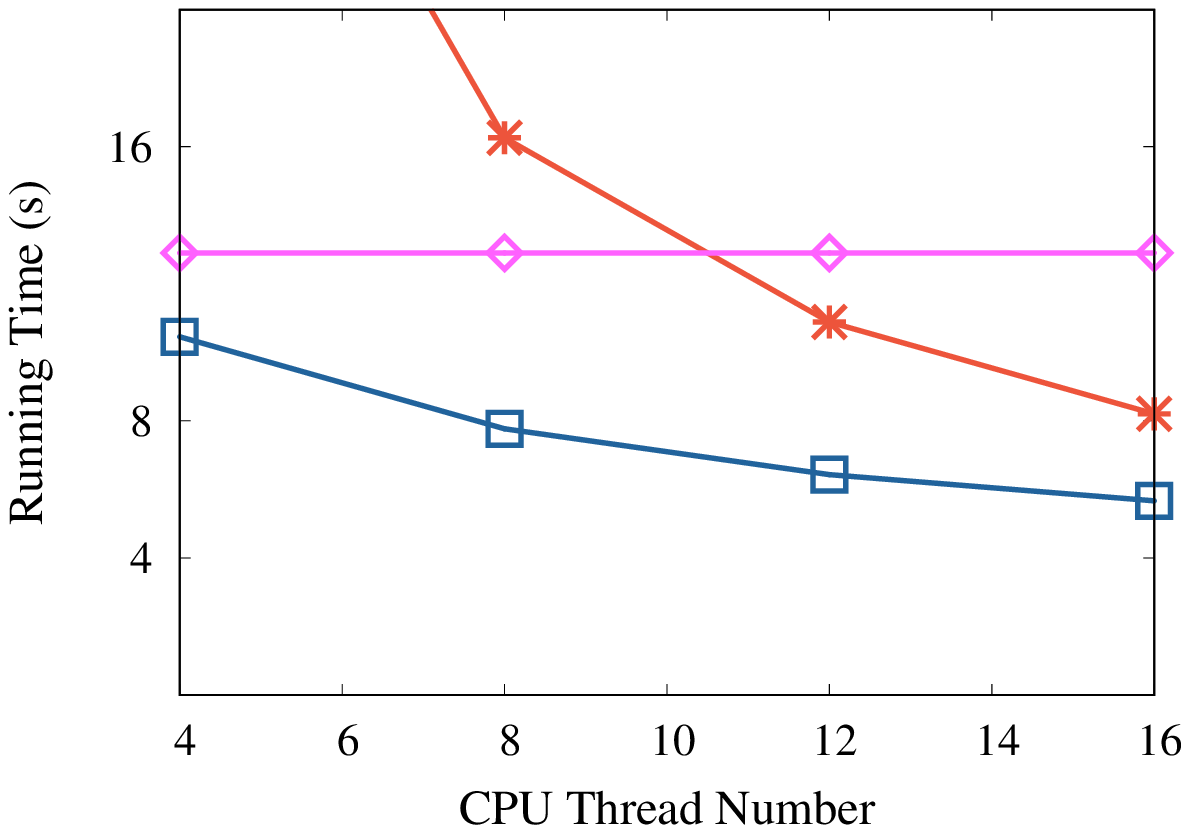}
}
\hspace{-1em}
\subfigure[\nf]{
  \includegraphics[width=0.5\columnwidth]{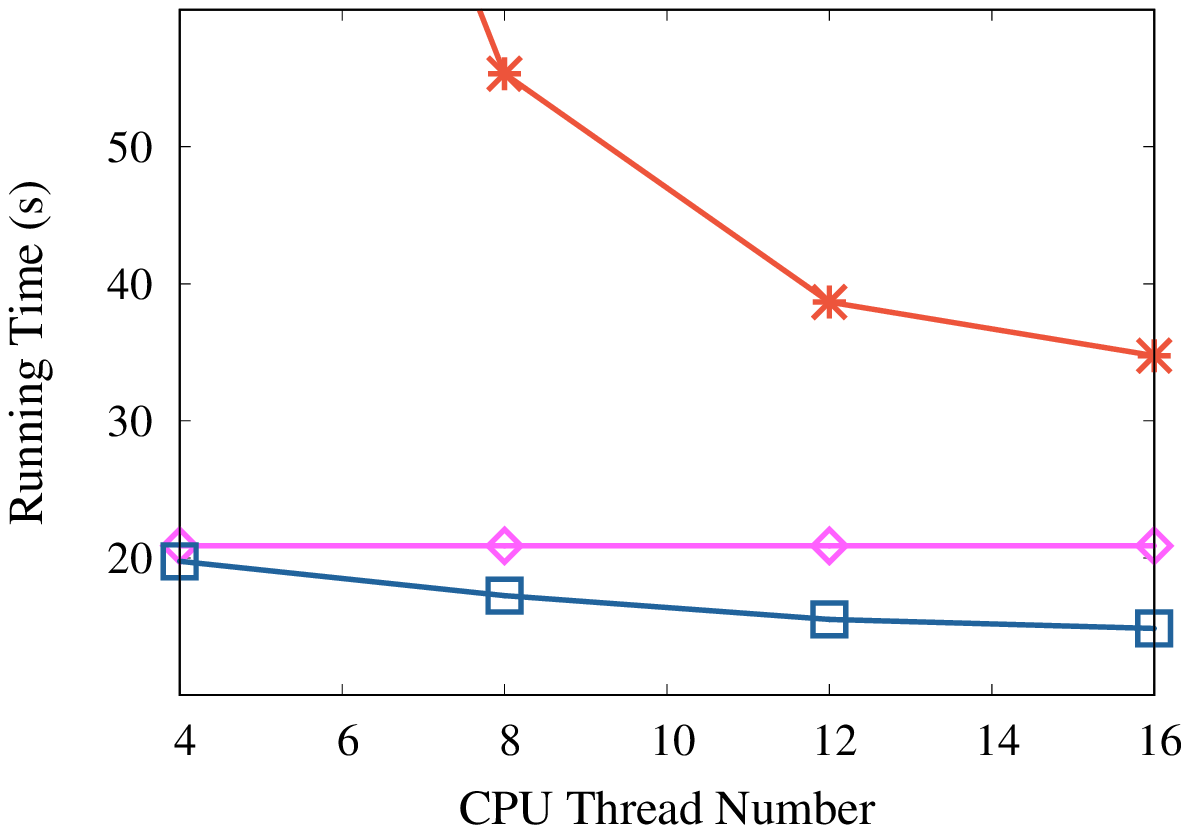}
}
\\
\hspace{-1em}
\subfigure[\ro]{
  \includegraphics[width=0.5\columnwidth]{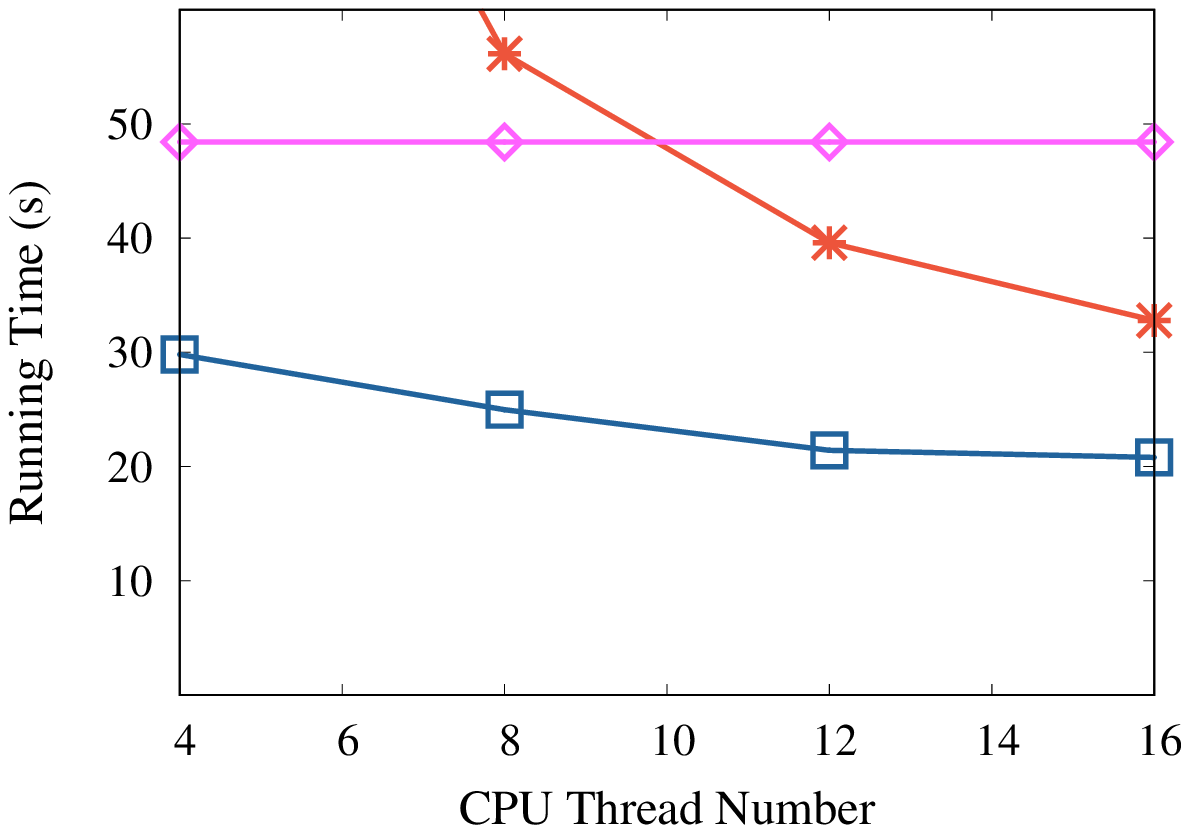}
}
\hspace{-1em}
\subfigure[\yh]{
  \includegraphics[width=0.5\columnwidth]{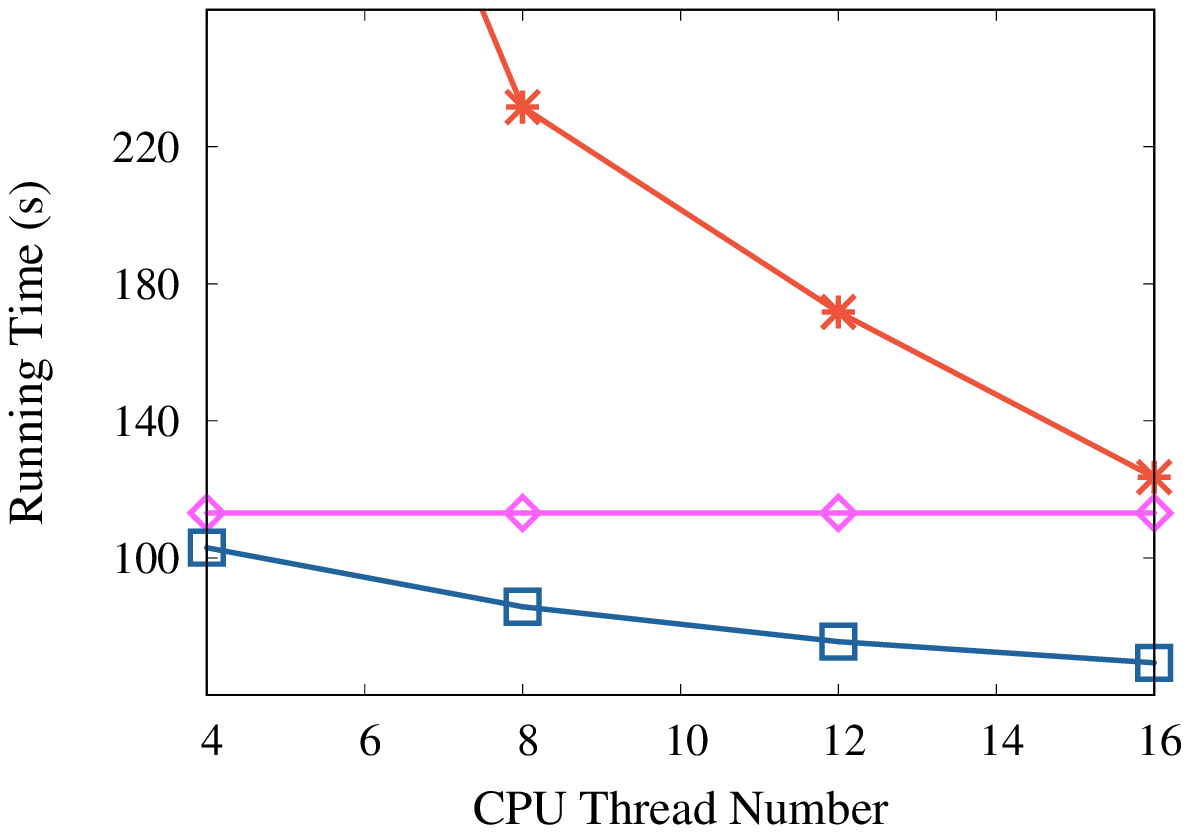}
}

\end{tabular}
\end{center}
\vspace{-1em}
\caption{Varying CPU Threads}
\label{fig:exp:cpu}
\end{figure}

In this experiment, we evaluate the adaptiveness of our algorithm by varying the CPU thread number from 4 to 16. The GPU parallel workers are fixed to the default value $128$. The running times of algorithms for different CPU thread numbers are reported in \reffig{exp:cpu}.

In contrast to \reffig{exp:gpu}, the running time of \alggpu is consistent, and the running time of \algcpu decreases when we use more CPU threads. \algopt is the fast algorithm on all settings and all datasets. For example, in \ro when the CPU thread number is 4, \algopt takes $29$ seconds, while \algcpu takes $109$ seconds, and \alggpu takes $48$ seconds. When the CPU thread number increases to 16, \algopt takes only $20$ seconds while \algcpu takes $33$ seconds. The decrease of the time of \algopt shows that our algorithm is adaptive to different CPU settings and can fully utilize the increasing computing power of CPUs.

\reffig{exp:cpu} and \reffig{exp:gpu} show the high efficiency of \algopt. When the gap between the computing power of CPU and GPU is limited (default setting), \algopt achieves a 1.4-2.3x speedup over \algcpu and a 1.4-2.3x speedup over \alggpu on all datasets. 
%
%
The experiments also show that the overhead cost of \algopt is minor. When the gap between the computing power of CPU and GPU is large, e.g., CPU uses 16 threads and GPU uses 512 parallel workers in \reffig{exp:gpu}, \algopt still achieves a slight speedup over \alggpu.

\subsection{Training Quality}
\label{subsec:exp:training}

In this experiment, we report the derived loss values (RMSE) of our algorithm \algopt during the training process to show the effectiveness of our method. The experiment will demonstrate the loss value of our algorithm finally converges to a reasonable value. \algcpu and \alggpu are also compared as references.

\begin{figure}[t!]
\begin{center}
\begin{tabular}[t]{c}
\subfigure{
  \includegraphics[width=0.9\columnwidth]{exp_dw/key.eps}
}
\vspace{-0.5em}
\\
\setcounter{subfigure}{0}
\hspace{-1em}
\subfigure[\ml]{
  \includegraphics[width=0.5\columnwidth]{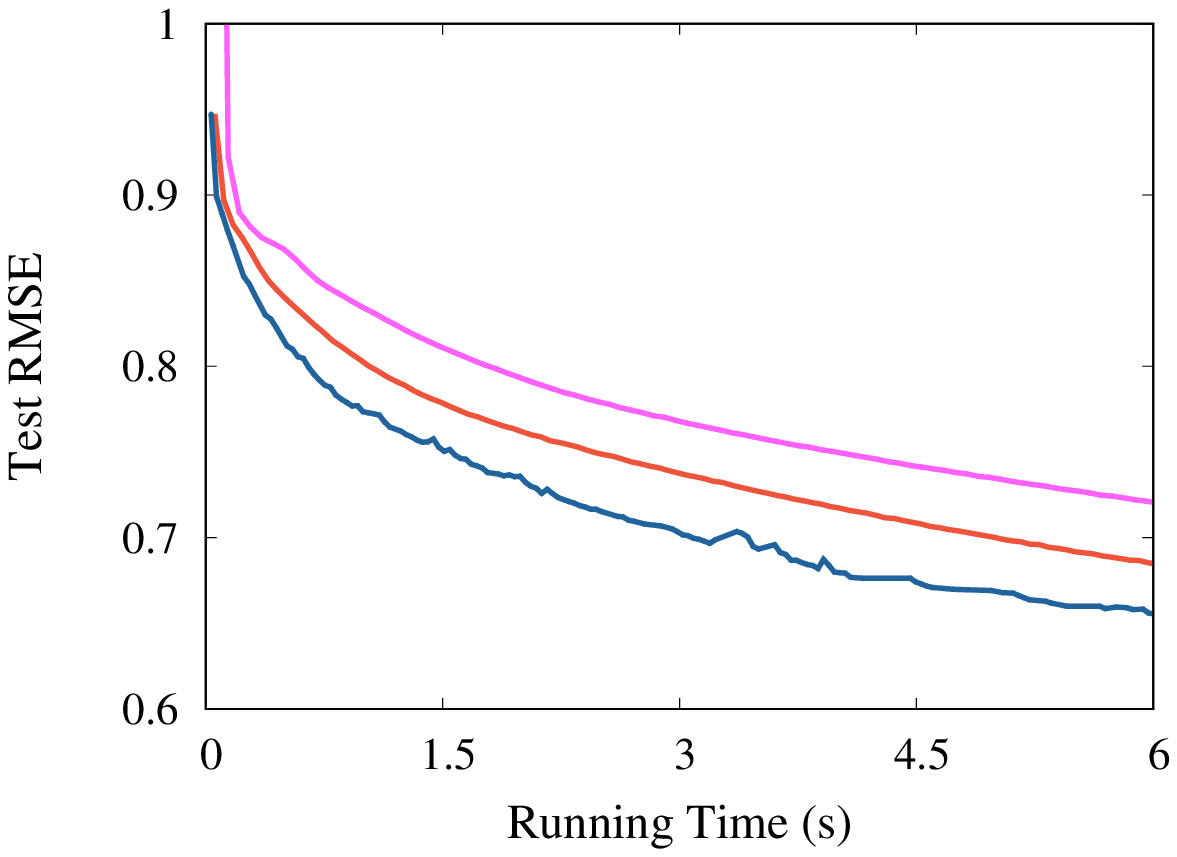}
}
\hspace{-1em}
\subfigure[\nf]{
  \includegraphics[width=0.5\columnwidth]{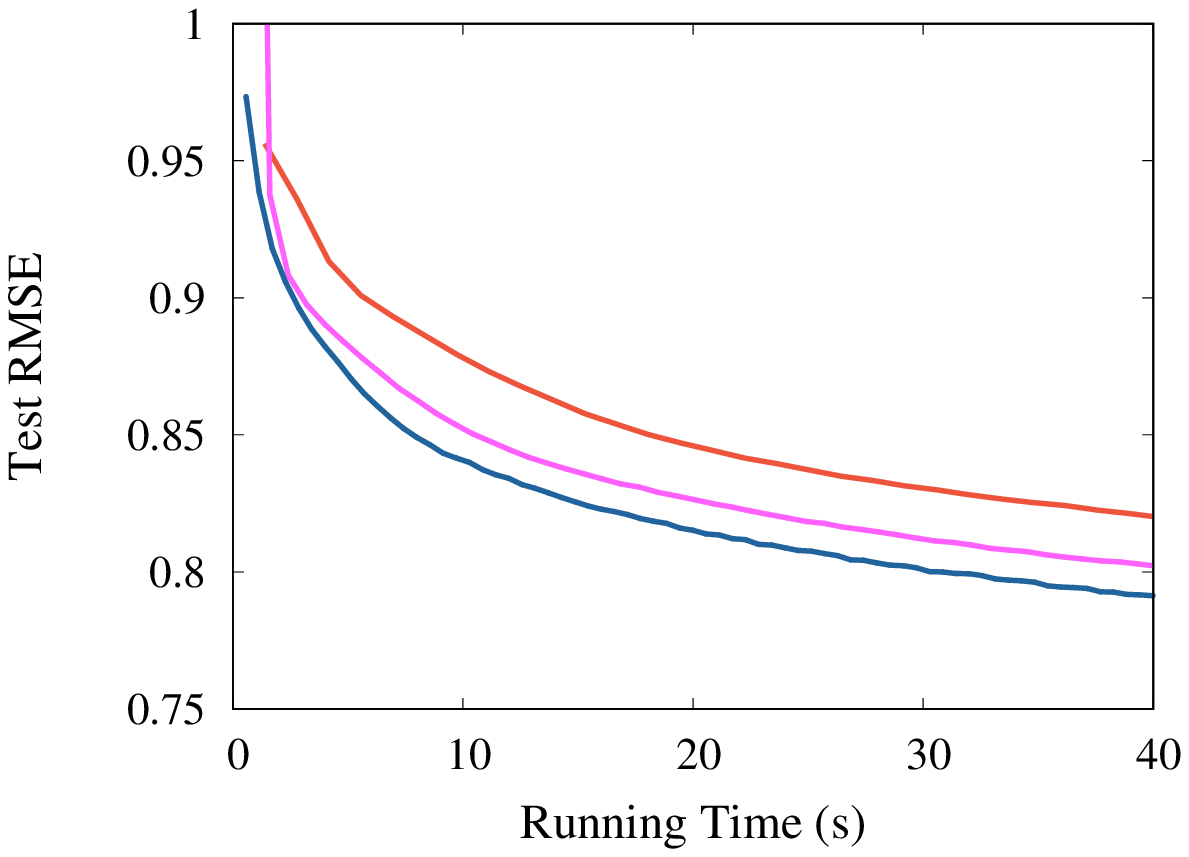}
}
\\
\hspace{-1em}
\subfigure[\ro]{
  \includegraphics[width=0.5\columnwidth]{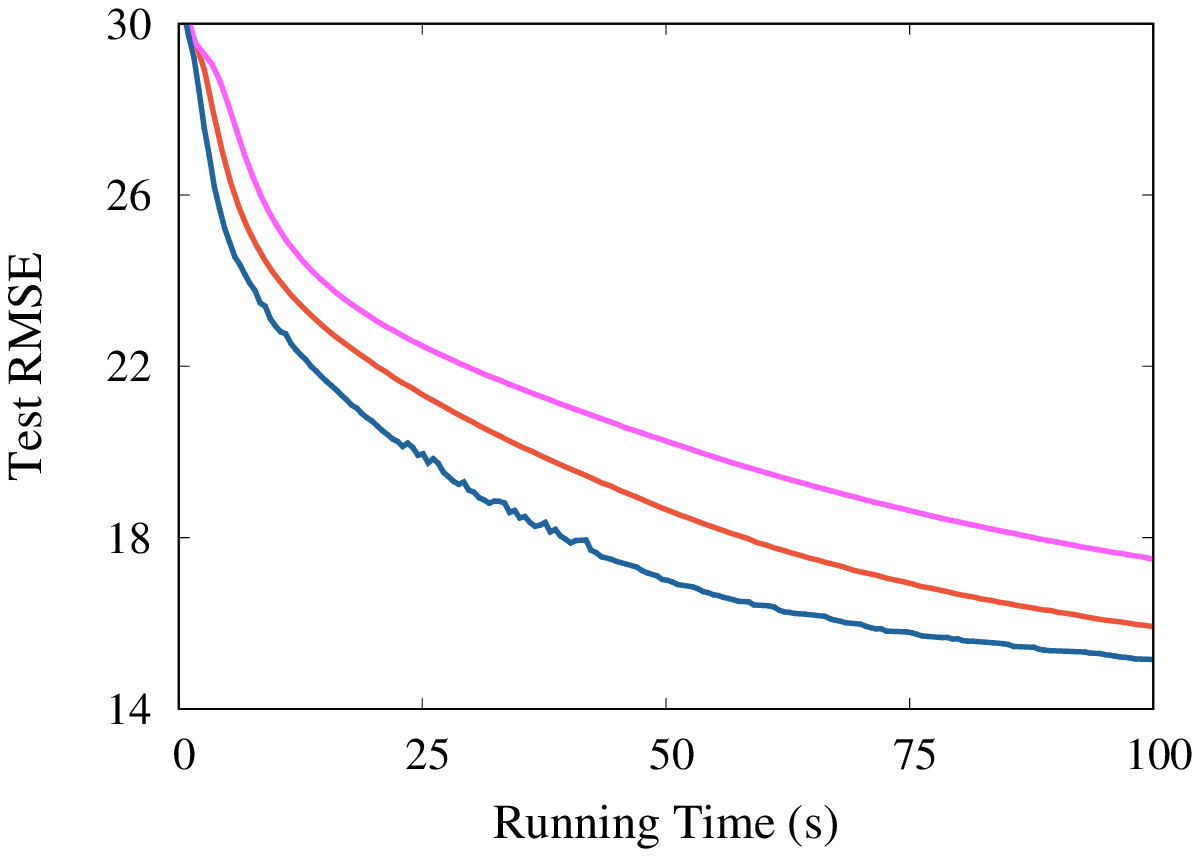}
}
\hspace{-1em}
\subfigure[\yh]{
  \includegraphics[width=0.5\columnwidth]{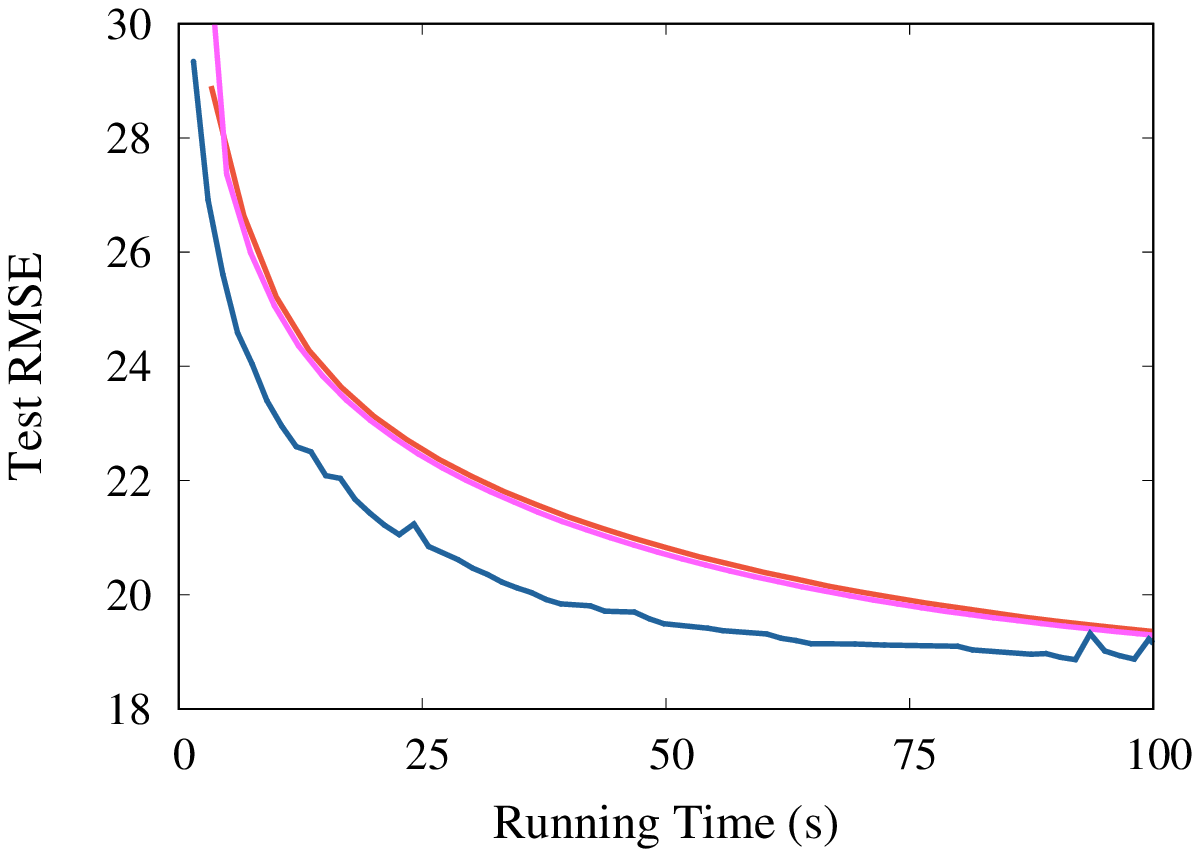}
}

\end{tabular}
\end{center}
\vspace{-1em}
\caption{Test RMSE over training time on four datasets}
\label{fig:exp:loss}
\end{figure}

The results are shown in \reffig{exp:loss}. The downward trends of \algopt are obvious, and the loss of \algopt converges in the shortest time. In addition, \algopt achieves a similar converged loss value compared with other algorithms, which shows the effectiveness of our algorithm. For example, in \yh, the loss of \algopt drops to $23$ in 10 seconds. At the same time, the loss values of \algcpu and \alggpu are $25.2$ and $25$, respectively. When time increases to 25 seconds, the loss of \algopt, \algcpu, and \alggpu drops to $20$, $22.5$, and $22.3$, respectively. Finally, all loss values of \algopt, \algcpu, and \alggpu stay about $19$.

\subsection{Matrix Division Strategy}
\label{subsec:exp:opt}

\begin{figure}[t!]
\begin{center}
\begin{tabular}[t]{c}
\subfigure{
  \includegraphics[width=0.9\columnwidth]{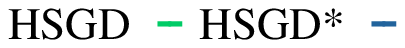}
}
\vspace{-0.5em}
\\
\setcounter{subfigure}{0}
\hspace{-1em}
\subfigure[\ml]{
  \includegraphics[width=0.5\columnwidth]{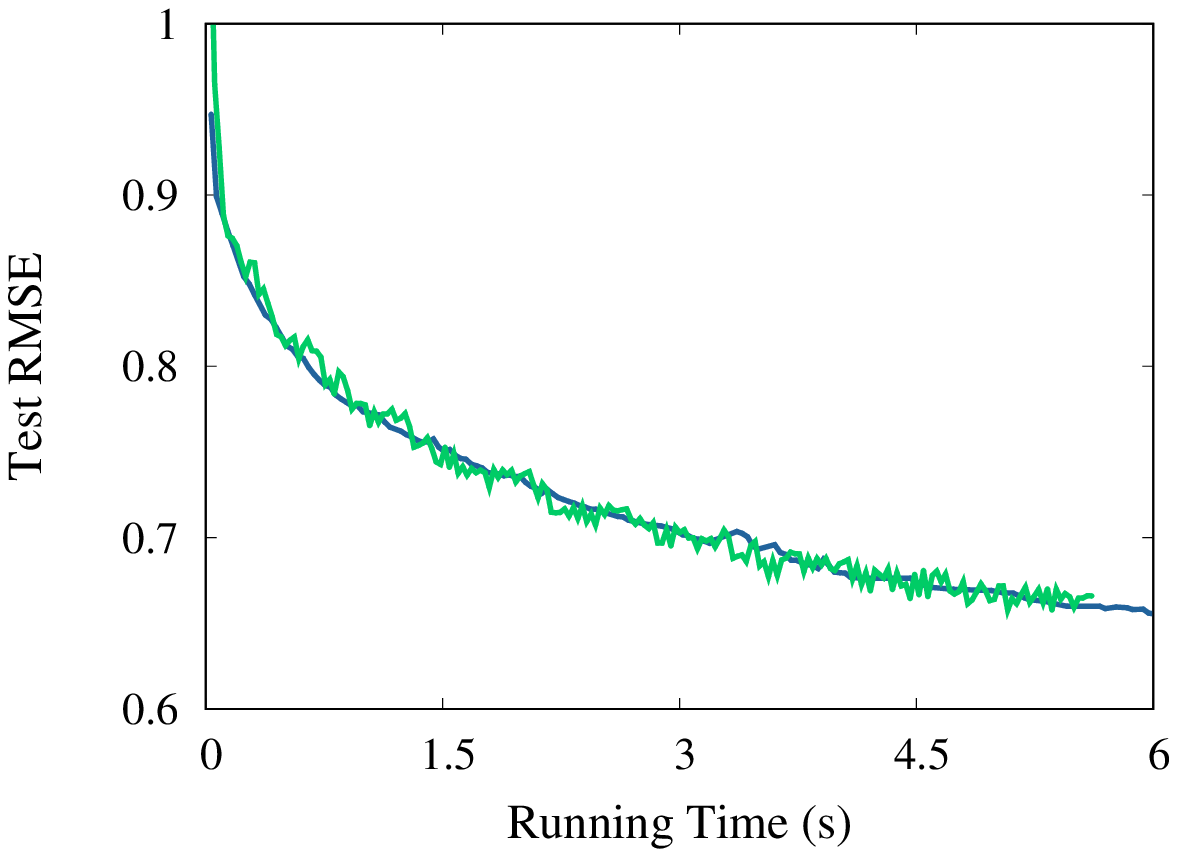}
}
\hspace{-1em}
\subfigure[\nf]{
  \includegraphics[width=0.5\columnwidth]{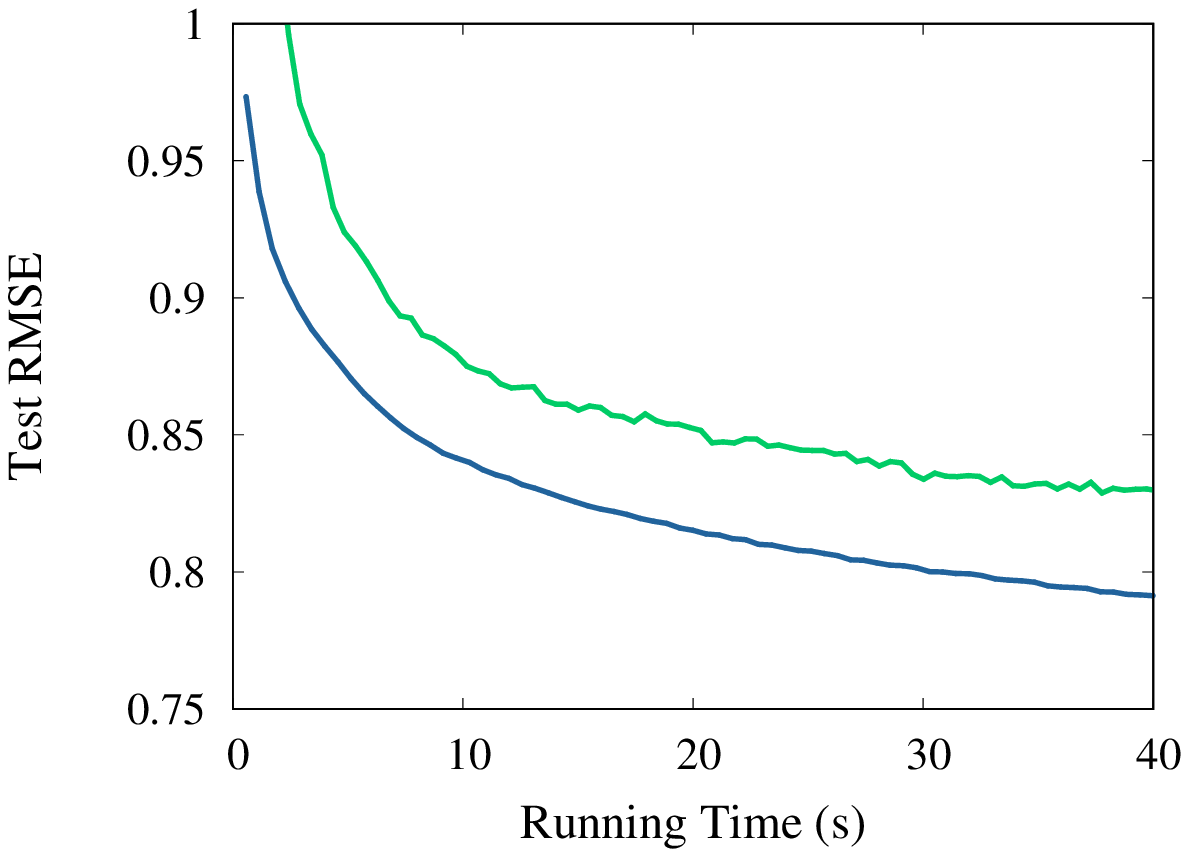}
}
\\
\hspace{-1em}
\subfigure[\ro]{
  \includegraphics[width=0.5\columnwidth]{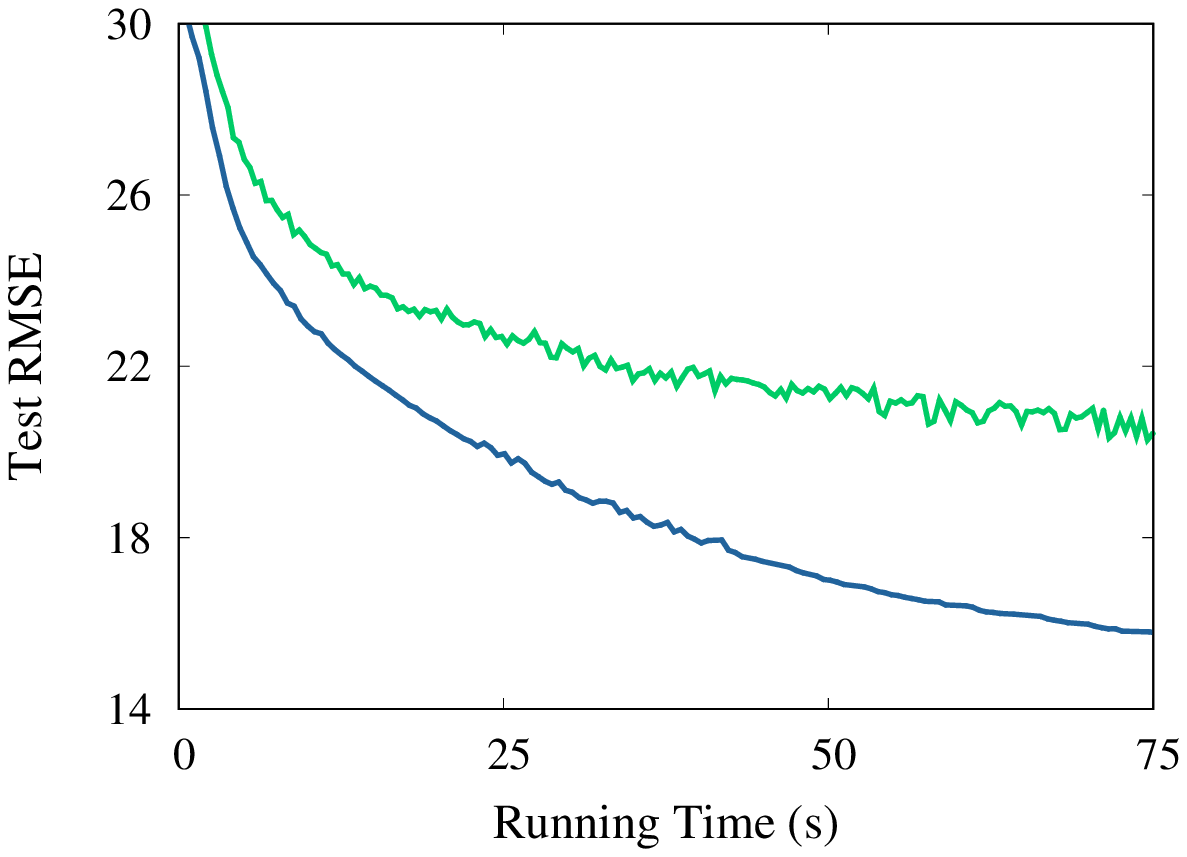}
}
\hspace{-1em}
\subfigure[\yh]{
  \includegraphics[width=0.5\columnwidth]{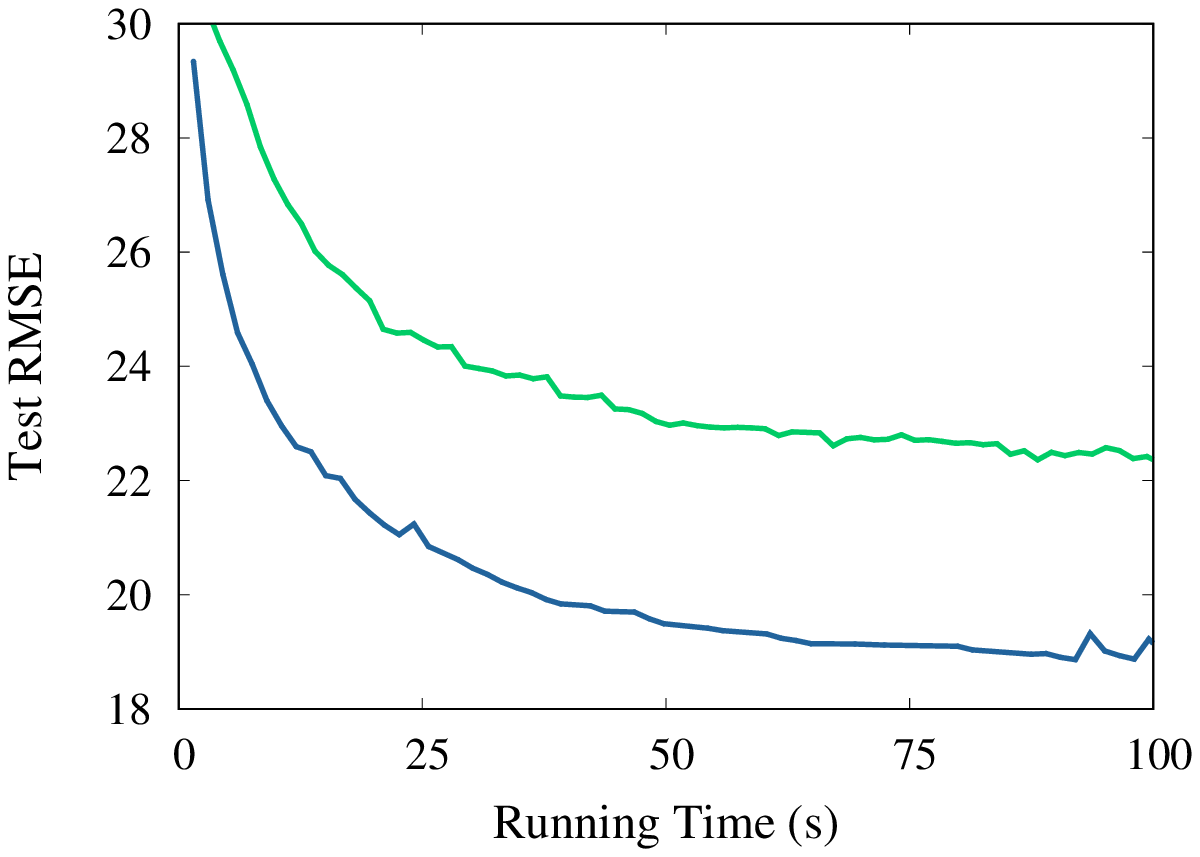}
}

\end{tabular}
\end{center}
\vspace{-1em}
\caption{Test RMSE over training time}
\label{fig:exp:convergence}
\end{figure}

We evaluate the effectiveness of our matrix division strategy in this experiment. For each dataset, we record the loss (RMSE) value on different running times of \algopt with \algnaive as a comparison. The result is shown in \reffig{exp:convergence}. 

We can see that the training quality of \algnaive is poor especially when processing relatively large datasets. This phenomenon is consistent with the discussion in \refex{poor_training_result}. The nonuniform matrix division in \algopt fixes this issue. Given the same running time, \algopt derives a smaller loss value than \algnaive, and the advantage of \algopt is obvious especially in large datasets. For example, given 50 seconds in \ro, the RMSE value for \algopt reaches to $17$, while that for \algnaive is only $21$. The result proves that nonuniform matrix division can utilize GPU resources better.

\vspace{-1em}
\begin{table}[ht!]
\scriptsize
\caption{Comparison of cost models}
\label{tab:exp:model}
\vspace{-0.5em}
\begin{tabular}{c|c|c|c|c|c}
\hline
\multicolumn{2}{c|}{Datasets} & \ml & \nf & \ro & \yh \\ \hline\hline
\multicolumn{6}{c}{Workload proportion} \\ \hline
\multirow{2}{*}{\algopt-Q} & C & 49.56\% & 55.98\% & 56.07\% & 56.46\% \\ \cline{2-6} 
 & G & 50.44\% & 44.02\% & 43.93\% & 43.54\% \\ \hline
\multirow{2}{*}{\algopt-M} & C & 55.91\% & 49.02\% & 49.75\% & 53.61\% \\ \cline{2-6} 
 & G & 44.09\% & 50.98\% & 50.25\% & 46.39\% \\ \hline\hline
\multicolumn{6}{c}{Running time} \\ \hline
\multicolumn{2}{c|}{\algopt-Q} & 0.92 s & 15.87 s & 13.07 s & 40.88 s \\ \hline
\multicolumn{2}{c|}{\algopt-M} & \textbf{0.89} s & \textbf{13.02} s & \textbf{12.08} s & \textbf{35.41} s \\ \hline
\end{tabular}
\vspace{-1em}
\end{table}

\subsection{Workload Balance}
\label{subsec:exp:balance}

We evaluate the effectiveness of techniques used to balance workloads in this section.

\subsubsection{Cost Models}


To show the effectiveness of our cost models, we report the proportion of workloads derived by our cost models with \cite{luk2009qilin} as a comparison in \reftab{exp:model}. To clearly reflect the algorithmic efficiency based on two cost models, we make these two methods run the same number of iterations, which is 20 in this experiment.

In \reftab{exp:model}, \algopt-Q represents the algorithm \algopt which uses Qilin \cite{luk2009qilin} to evaluate the working efficiency of hardware. \algopt-M represents the algorithm \algopt which uses our model in \refsec{model} to evaluate the working efficiency of hardware. Note that for fairness, both \algopt-Q and \algopt-M do not include the dynamic scheduling strategy in \refsec{task} to further balance workloads. "C" and "G" in the table represent the assigned proportion of workloads to CPUs and GPUs, respectively.

The practical running times of \algopt-Q and \algopt-M are also reported. The running time of \algopt-M is smaller than that of \algopt-Q on all datasets. This result proves that our cost model can derive a more accurate estimation for the working efficiency of hardware. We can find that \algopt-M prefers to assign more work to GPU compared with \algopt-Q on all datasets except \ml. 
For \ml, \algopt-M observes that the performance of GPU is not strong when processing a small dataset (\refob{small_gpu}). Therefore, it assigns less work to GPU. The effectiveness of \algopt-M becomes considerable when the dataset is large. Note that given a smaller target loss, both algorithms will require more iterations, and the advantage of our method will become obvious. For example, to achieve the predefined loss value of Yahoo!Music in \refsubsec{exp:overall}, \algopt needs 46 iterations, which is more than twice that of this experiment.

\vspace{-1em}
\begin{table}[ht!]
\centering
\small
\caption{Effectiveness of dynamic scheduling}
\label{tab:exp:dynamic}
\vspace{-0.5em}
\begin{tabular}{c|c|c}
\hline
Dataset     & \algopt-M & \algopt \\ \hline
\ml   & 0.89 s      & \textbf{0.84} s  \\ \hline
\nf   & 13.02 s     & \textbf{11.42} s  \\ \hline
\ro   & 12.08 s     & \textbf{10.58} s    \\ \hline
\yh   & 35.41 s     & \textbf{30.96} s   \\ \hline
\end{tabular}
\end{table}

\subsubsection{Dynamic Scheduling}

We evaluate the effectiveness of the dynamic scheduling strategy (\refsec{task}). Similar to the experiment for cost models, we use \algopt-M to denote our final algorithm without the dynamic scheduling technique to further balance workloads. The running times of \algopt-M and \algopt on all datasets are shown in \reftab{exp:dynamic}. 
%

The result shows \algopt is faster than \algopt-M on all datasets. Note that in \ml with relatively small size, the computing power of GPU cannot be saturated, which degrades the effectiveness of the dynamic scheduling. As a result, the improvement of dynamic scheduling on \ml is minor.
By combining the new cost models and the dynamic scheduling strategy, our final algorithm achieves a significant improvement in balancing workloads of MF.

\section{Conclusion}
\label{sec:conclusion}

We propose an efficient parallel MF algorithm on heterogeneous CPU-GPU systems. To fully utilizes the processing power of GPUs, our approach divides the input matrix using a nonuniform method and assigns large blocks to GPUs. We design a new cost model to estimate the performance of computing resources. We balance workloads by combining the cost models and dynamic scheduling in practice. 

\balance
\bibliographystyle{IEEEtran}
\bibliography{scRef}

\end{document}